\documentclass[a4paper,11pt]{article}
\pdfoutput=1 % if your are submitting a pdflatex (i.e. if you have
\usepackage{jcappub} % for details on the use of the package, please
\usepackage[T1]{fontenc} % if needed

\usepackage{amsmath}
\usepackage{graphicx,epsf,hyperref}

\usepackage{color}
\usepackage[dvipsnames]{xcolor}

\newcommand{\n}{\nabla}
\newcommand{\p}{\partial}

\newcommand{\bfk}{\mbox{\boldmath$k$}}

\newcommand{\bfq}{\mbox{\boldmath$q$}}

\newcommand{\bfn}{\mbox{\boldmath$n$}}
\newcommand{\de}{\mathrm{d}}

\def\be{\begin{equation}}
\def\ee{\end{equation}}

\title{\textbf Cosmology with the EFTofLSS and BOSS: dark energy constraints and a note on priors}

\author[a]{Pedro Carrilho,}
\author[a,b]{Chiara Moretti}
\author[a,c]{and Alkistis Pourtsidou}

\affiliation[a]{Institute for Astronomy, The University of Edinburgh, Royal Observatory, Edinburgh EH9 3HJ, UK}
\affiliation[b]{INAF – Osservatorio Astronomico di Trieste, Via Tiepolo 11, I-34143 - Trieste, Italy}
\affiliation[c]{Higgs Centre for Theoretical Physics, School of Physics and Astronomy,
The University of Edinburgh, Edinburgh EH9 3FD, UK}

\emailAdd{pedro.carrilho@ed.ac.uk}
\emailAdd{cmoretti@ed.ac.uk}
\emailAdd{alkistis.pourtsidou@ed.ac.uk}

\abstract{

\noindent We analyse the BOSS DR12 multipoles of the galaxy power spectrum jointly with measurements of the BAO scale for three different models of dark energy. We use recent measurements performed with a windowless estimator, and an independent and fast pipeline based on EFTofLSS modelling implemented via the \texttt{FAST-PT} algorithm to compute the integrals of the redshift-space loop corrections. We accelerate 
our analysis further by using the \texttt{bacco} linear power spectrum emulator instead of a Boltzmann solver. We perform two sets of analyses: one including $3\sigma$ Planck priors on $A_s$ and $n_s$, and another that is fully CMB-free, i.e., letting the primordial parameters vary freely. The first model we study is $\Lambda$CDM, within which we reproduce previous results obtained with the same estimator. We find a low value of the scalar amplitude in the CMB-free case, in agreement with many previous EFT-based full-shape analyses of the BOSS data. We then study $w$CDM, finding a lower value of the amplitude in the CMB-free run, coupled with a preference for phantom dark energy with $w=-1.17^{+0.12}_{-0.11}$, again in broad agreement with previous results. Finally, we investigate the dark scattering model of interacting dark energy, which we label $wA$CDM. In the CMB-free analysis, we find a large degeneracy between the interaction strength $A$ and the amplitude $A_s$, hampering measurements of those parameters. On the contrary, in our run with a CMB prior, we are able to constrain the dark energy parameters to be $w=-0.972^{+0.036}_{-0.029}$ and $A = 3.9^{+3.2}_{-3.7}$, which show a 1$\sigma$ hint of interacting dark energy. This is the first measurement of this parameter and demonstrates the ability of this model to alleviate the $\sigma_8$ tension. Our analysis can be used as a guide for the analysis of any model with scale-independent growth. Finally, we study the dependence of the results on the priors imposed on the nuisance parameters and find these priors to be informative, with their broadening generating shifts in the contours. We argue for an in depth study of this issue, which can affect current and forthcoming analyses of LSS data.
}

\begin{document}
\maketitle
\flushbottom

\section{Introduction}
\label{sec:intro}

The standard model of cosmology has enjoyed great success in describing the observations of cosmological surveys, from the cosmic microwave background (CMB)~\cite{Aghanim:2018eyx} to the large-scale structure (LSS)~\cite{Alam:2020sor}. In spite of these successes, the main ingredients of this $\Lambda$CDM model -- cold dark matter (CDM) and a cosmological constant ($\Lambda$) -- are still poorly understood from the fundamental point of view and it is possible that their replacement with alternative components is necessary. Alternative theories for dark energy include modified gravity, as well as quintessence, with the possibility of additional interactions with dark matter~\cite{Caldwell:1997ii,Amendola:1999er,Peebles:2002gy,Copeland:2006wr,Nojiri:2006ri,Sotiriou:2008rp, DeFelice:2010aj,Clifton:2011jh,Bertolami:2012xn, Pourtsidou:2013nha}.

In addition to its theoretical issues, the $\Lambda$CDM model has recently been challenged by a number of observed tensions between different sets of cosmological data~\cite{Verde:2019ivm,Knox:2019rjx,Jedamzik_2021,Di_Valentino_2021,perivolaropoulos2021challenges}. 
A particularly interesting discrepancy is that between the amplitude of perturbations as observed at late times, when compared to the estimate from CMB observations. This discordance arises in the determination of $\sigma_8$, the variance of density fluctuations averaged over spheres of $8~{\rm Mpc}/h$, which is measured to be lower than expected from the Planck best-fit $\Lambda$CDM cosmology~\cite{Aghanim:2018eyx}. This has been seen in multiple probes of the growth of structure, such as in measurements of weak lensing and photometric clustering~\cite{Heymans:2013fya,HSC:2018mrq,Abbott:2021bzy,KiDS:2020suj,DES:2021bvc,Secco:2021vhm,2021A&A...646A.140H,Amon:2022ycy}, spectroscopic clustering~\cite{Blake:2011rj,Reid:2012sw,Macaulay_2013,Beutler:2013yhm,Beutler_2016,Gil-Marin:2016wya,DAmico:2019fhj,Philcox:2021main,Ivanov:2019pdj,Troster:2019ean} and galaxy clusters~\cite{deHaan:2016qvy}. In particular, the values of the best-constrained quantity, $S_8\equiv\sigma_8\sqrt{\Omega_m/0.3}$, are found to be up to 10\% lower ($S_8\approx0.77$) than the value inferred by Planck ($S_8\approx0.83$).

A number of models have been proposed to alleviate this tension, with dark sector interactions being a promising possibility~\cite{Amendola:1999er,Farrar:2003uw}. The main reason for this is that the addition of the dark interaction can slow the growth of dark matter density fluctuations, thus reducing their amplitude at late times. Different flavours of these interacting dark energy (IDE) models have been explored, most of which include energy exchange as a result of the interaction. While these energy-transfer models have had some recent success in fitting observations~\cite{Nunes:2022bhn,Barros:2018efl,vandeBruck:2019vzd}, they typically are heavily constrained by CMB data~\cite{Bean_2008,Xia:2009zzb,Amendola_2012,Gomez-Valent:2020mqn} due to the substantial modifications they introduce to background evolution. Here we focus instead on models for which only momentum is exchanged between dark matter and dark energy~\cite{Simpson:2010vh,Pourtsidou:2013nha,Skordis:2015yra,Richarte:2014yva}. This avoids any modification to background evolution beyond that generated by dynamic dark energy, which allows for better fits to data~\cite{Pourtsidou:2016ico,Linton:2017ged,Linton:2021cgd,ManciniSpurio:2021jvx}. In this work we investigate the Dark Scattering model~\cite{Simpson:2010vh}, for which dark matter particles are hypothesised to scatter elastically with dark energy, exchanging only momentum. This is a well-developed model, thoroughly studied with N-body simulations~\cite{Baldi:2014ica,Baldi:2016zom} and with robust nonlinear modelling already developed~\cite{Carrilho:2021hly,Carrilho:2021rqo}. Similarly to other interacting models, it is able to alleviate the $S_8$ tension by slowing down the clustering of dark matter~\cite{Bose:2017jjx,Baldi:2014ica,Baldi:2016zom,Linton:2021cgd}. Similar models which also have this property include Refs.~\cite{Lesgourgues:2015wza,Linton:2017ged,Buen-Abad:2017gxg,Kase:2019mox,Chamings:2019kcl,Amendola:2020ldb,1869592,Ferlito:2022mok,Vagnozzi:2019kvw}.

Current and forthcoming stage IV galaxy surveys, such as the Dark Energy Spectroscopic Instrument (DESI)~\cite{Aghamousa:2016zmz}, ESA's Euclid satellite mission~\cite{Laureijs:2011gra,Blanchard:2019oqi}, Rubin's Legacy Survey of Space and Time (LSST)~\cite{Mandelbaum:2018ouv}, and the Nancy Grace Roman space telescope~\cite{spergel2015wide} are expected to pin down the details of the dark sector to unprecedented accuracy and will allow for the distinction between many alternatives to $\Lambda$CDM, including IDE models. To ensure the success of these efforts, it is important that the theoretical predictions from these non-standard cosmologies have the highest quality, as the volume of data and its precision will be unprecedented, leaving no place for theoretical systematics to hide~\cite{Chisari:2019tus, Markovic:2019sva, Bose:2019psj, Schneider:2019snl, Nishimichi:2020tvu, Martinelli:2020yto, Pezzotta:2021vfn, Secco:2021vhm,DES:2021bvc}. This has led to the development of new techniques extending the range of scales for which predictions are trustworthy. A key recent development is the effective field theory of large-scale structure (EFTofLSS)~\cite{Baumann:2010tm,Carrasco:2012cv}. This technique models the unknown effects of very small scales on mildly nonlinear scales via the introduction of effective stresses and adds corresponding time-dependent parameters, improving the fits to simulations~\cite{Nishimichi:2020tvu}.

These new techniques are powerful enough to be useful already in the analysis of current data. In particular, the EFTofLSS has been used by different groups to re-analyse the summary statistics measured by BOSS, both in the context of $\Lambda$CDM~\cite{DAmico:2019fhj, Ivanov:2019pdj, Troster:2019ean, Philcox:2021main,DAmico:2022osl}, as well as for extended models~\cite{DAmico:2020kxu,Chudaykin:2020ghx,Nunes:2022bhn}. Our aim in this paper is exactly this -- use the EFTofLSS to analyse the BOSS data in order to constrain exotic dark energy and dark scattering interactions. We use an independent implementation of the EFTofLSS-based modelling validated in previous work~\cite{Carrilho:2021hly} to predict the multipoles of the power spectrum and run an MCMC analysis for all cosmological parameters on the BOSS DR12 data, including also post-reconstructed BAO data from that release, as well as additional pre-reconstruction measurements of the BAO scale in a broad redshift range. We are thus able to constrain the strength of the interaction, in parallel with the dark energy equation of state (EOS) parameter, for the first time.

This paper is organised as follows. The data and the methods we use in our analysis pipeline are described in Section~\ref{sec:data}. Our baseline results are shown in Section~\ref{sec:resultsmain}, with full contour plots shown in Appendix~\ref{sec:full_cont} and a detailed discussion on the information content of the priors of nuisance parameters and their effects on the results in Appendix~\ref{sec:prior}. Finally, we detail our conclusions in Section~\ref{sec:conclusions}.

\section{Data and Methods}
\label{sec:data}

\subsection{Datasets}

The main dataset we analyse is the Baryon Oscillation Spectroscopic Survey (BOSS) Data Release 12 (DR12) measurement of the power spectrum multipoles~\cite{SDSS:2011jap,2015ApJS..219...12A,Gil-Marin:2015sqa,Beutler_2016}. This is based on the CMASS and LOWZ samples, containing in excess of a million galaxies, observed in the Northern and Southern galactic caps (NGC and SGC). These are separated into two redshift bins (0.2 < z < 0.5 and 0.5 < z < 0.75) with effective redshifts $z_1=0.38$ and $z_3=0.61$. This results in four distinct sets of measurements with differing selection criteria, number density and volume. 

We use here the measurements of the windowless power spectrum multipoles, as provided by Ref.~\cite{Philcox:2021main}\footnote{The measured spectra and the code to generate them can be found at \url{github.com/oliverphilcox/BOSS-Without-Windows}, while a likelihood using them can be found at \url{github.com/oliverphilcox/full_shape_likelihoods}.} and based on the estimator of Ref.~\cite{Philcox:2020vbm} (see also Ref.~\cite{Philcox:2021ukg} for a similar treatment for the bispectrum). This allows our theoretical predictions to be directly compared to data, without the need for a convolution with the window function. Additionally, these measurements do not suffer from the normalisation issue discussed in Refs.~\cite{deMattia:2019vdg,deMattia:2020fkb,Beutler:2021eqq}, since, as described in Ref.~\cite{Philcox:2021main}, this estimator does not require the 2-point function of the mask to be computed explicitly, which is where the issue would arise.

Additionally, we include the post-reconstructed BAO measurements from the same BOSS data release as provided in Ref.~\cite{Philcox:2021main}, giving 4 different measurements of the parameters $\alpha_\parallel$, $\alpha_\perp$ at $z=0.38$ and $z=0.61$. We also use additional pre-reconstruction measurements of the BAO scale, including low redshift measurements of the volume distance from 6DF at $z=0.106$~\cite{Beutler:2011hx} and from the SDSS DR7 MGS at $z=0.15$~\cite{Ross:2014qpa}, as well as high redshift measurements of the Hubble and angular diameter distance from Lyman-$\alpha$ forest auto-correlation and cross-correlation with quasars from eBOSS DR14 measurements at $z=2.334$~\cite{duMasdesBourboux:2020pck}.

The covariance matrix for the full shape measurements is also provided by Ref.~\cite{Philcox:2021main} and includes the cross-covariance between the BAO measurements and the multipoles of the power spectrum. It was produced using the publicly available suite of 2048 ‘MultiDark-Patchy’ mock catalogs~\cite{Kitaura:2015uqa,Rodriguez-Torres:2015vqa}.

We include a Gaussian prior on the baryon density from Big Bang Nucleosythesis, $\omega_b=0.02268\pm0.00038$ ~\cite{Ivanov:2019pdj,Aver:2015iza,Cooke:2017cwo,Adelberger:2010qa,Pisanti:2007hk}. For each case, we perform two different analysis, one of which includes a $3\sigma$ Planck prior on the parameters of the primordial power spectrum, using $\log A_s=3.044\pm 0.042$ and $n_s=0.9649\pm0.012$~\cite{Aghanim:2018eyx}. This prior on $A_s$ is somewhat unjustified in the context of $\Lambda$CDM, since $f\sigma_8$ is well measured by RSDs, with $f$ being simply related to $\Omega_m(z)$, which is in turn well measured by the BAO data; and $\sigma_8$ being directly connected to $A_s$ in that model. However, as discussed below in more detail, the interacting dark energy model studied here modifies growth and breaks the connection between the growth rate, $f$, and $\Omega_m(z)$ as well as the simple relation between $A_s$ and $\sigma_8$. For this reason, introducing a prior on $A_s$ is crucial for breaking degeneracies in the interacting dark energy model and we consider that our baseline analysis in that case. For completeness, we perform both analyses for all three dark energy models we consider, but our more general analysis is fully CMB-free, and therefore does not include any prior on the primordial parameters.

\subsection{Modelling}
\label{sec:model}

We consider three distinct dark energy models, starting with $\Lambda$CDM, then moving to $w$CDM and finally investigating the Dark Scattering model. Since these models are built as extensions of each other, we describe here the most general case of Dark Scattering, from which the other models can be easily obtained via appropriate limits. This is an interacting dark energy model, in which cold dark matter exchanges momentum with dark energy via elastic scattering. This model does not introduce an exchange of energy between those two species and for that reason does not change the expansion history with respect to the $w$CDM model. Here we briefly review the evolution of the different species as well as the modelling used to construct the galaxy power spectrum in redshift space.

\subsubsection{Evolution equations}

In the assumed flat Friedmann-Lema\^{i}tre-Robertson-Walker spacetime, the Friedmann equation at late times is given by 
\be
H^2=H_0^2\left(\Omega_{m}a^{-3}+\Omega_{\rm{DE}}a^{-3(1+w)}\right)\,,
\ee
where $H=\dot{a}/a$ is the Hubble rate, with $a$ the scale factor and $H_0$ its value at $z=0$; $\Omega_{m}$ is the matter density parameter at $z=0$, including cold dark matter ($\Omega_{c}$), baryons ($\Omega_{b}$) and massive neutrinos ($\Omega_{\nu}$); dark energy is assumed to be dynamic, with constant equation of state parameter $w\equiv P_{\rm DE}/\rho_{\rm DE}$ and density parameter $\Omega_{\rm{DE}}$.

We describe the matter fluctuations with perturbation theory~\cite{Malik:2008im} in the Newtonian approximation~\cite{Bernardeau:2001qr}. Within the Dark Scattering model~\cite{Simpson:2010vh,Baldi:2014ica,Baldi:2016zom, Carrilho:2021hly, Carrilho:2021rqo}, the only equations that are modified relative to $w$CDM are the momentum evolution equations for dark energy and dark matter. Under the assumption that the dark energy sound speed is unity, we can neglect dark energy fluctuations and their evolution. Given that, the evolution equations for dark matter fluctuations are given by
\begin{align}
    &a\p_a\delta(\mathbf{k})+\Theta(\mathbf{k})=-\int{\frac{\rm{d}^3\mathbf{k_1}\rm{d}^3\mathbf{k_2}}{(2\pi)^3}\delta^{(3)}(\mathbf{k}-\mathbf{k_1}-\mathbf{k_2})}\alpha(\mathbf{k_1},\mathbf{k_2})\Theta(\mathbf{k_1})\delta(\mathbf{k_2})\,,\label{dens_eq}\\
    &a\p_a\Theta(\mathbf{k})+\left(2+(1+w)\xi \frac{\rho_{\rm DE}}{H}+\frac{a\p_aH}{H}\right)\Theta(\mathbf{k})-\left(\frac{k}{aH}\right)^2\phi(\mathbf{k})=\nonumber\\
    &\quad\quad\quad\quad\quad\quad\quad-\frac12\int{\frac{\rm{d}^3\mathbf{k_1}\rm{d}^3\mathbf{k_2}}{(2\pi)^3}\delta^{(3)}(\mathbf{k}-\mathbf{k_1}-\mathbf{k_2})}\beta(\mathbf{k_1},\mathbf{k_2})\Theta(\mathbf{k_1})\Theta(\mathbf{k_2})\,,\label{vel_eq}
\end{align}
where $\delta$ is the density contrast, $\Theta$ is the normalized divergence of the velocity field, time derivatives $\p_a$ are written with respect to the scale factor and $\alpha$ and $\beta$ are the usual nonlinear couplings~\cite{Jain:1994,Bernardeau:2001qr,Carrilho:2021hly}. The key difference between this and the standard case is the interaction term, which depends on the interaction cross section, $\sigma_{\rm DS}$, and the dark matter mass, $m_c$, via $\xi=\sigma_{\rm DS}/m_c$.

These equations are solved in conjunction with the Poisson equation for the gravitational field order by order in perturbations, within the Einstein-de Sitter (EdS) approximation, up to 1-loop order. Additionally, we evolve the fluctuations of baryons and dark matter as a single species, using the effective coupling approximation employed in Ref.~\cite{Carrilho:2021rqo}. This approximation consists of using Eqs.~\eqref{dens_eq} and \eqref{vel_eq} for the evolution of all cold species, while taking into account the effect of non-interacting baryons by lowering the effective coupling between dark matter and dark energy. In practice, this is done by substituting the coupling $\xi$ with the effective coupling $\bar\xi$ defined by
\be
\bar\xi=\frac{f_c}{1+C_0(1-f_c)}\xi\,,
\label{eff_xi}
\ee
in which $f_c\equiv\Omega_c/\Omega_m$ is the dark matter fraction and $C_0=(1+w)\xi\rho_{\rm DE\,0}/H_0$ is the coupling term in Eq.~\eqref{vel_eq} evaluated at $z=0$. The effective strength of the interaction, and the parameter we will ultimately probe is therefore the combination
\be
A\equiv(1+w) \bar\xi\,.
\label{A_def}
\ee
We denote the Dark Scattering model in this work by $wA$CDM, to make it explicit that this is a simple extension to $w$CDM with all distinguishing effects parameterised by the interaction parameter $A$.

Finally, massive neutrinos are approximated to evolve linearly, with their effect taken into account via the linear power spectrum that will later be used in constructing the galaxy power spectrum. 

\subsubsection{EFTofLSS power spectrum}

To construct the galaxy power spectrum in redshift space, we use the EFTofLSS model of Refs.~\cite{Ivanov:2019pdj,Chudaykin:2020} (see also Refs.~\cite{Perko:2016puo, delaBella:2017qjy,DAmico:2019fhj, delaBella:2018fdb} for other similar models), which was also used in the context of Dark Scattering in Ref.~\cite{Carrilho:2021hly} and shown to be more effective in describing the power spectrum of biased tracers compared to alternatives such as the Taruya-Nishimichi-Saito (TNS) model~\cite{Taruya:2010mx}. The model we adopt includes a number of ingredients which we now briefly review.

The bias model used is based on a perturbative and gradient expansion of the galaxy density field (see Refs.~\cite{McDonald:2009dh,Assassi_2014,Desjacques:2016bnm,Abidi:2018eyd} for details), given by
\be
\delta_g = b_1 \delta_{cb} + \frac{b_2}{2} \delta_{cb}^2 + b_{\mathcal{G}_2} \mathcal{G}_2 + b_{\Gamma_3} \Gamma_3 +\epsilon,
\ee
where $b_1$ is the linear bias, $b_2$ is the quadratic bias and $\mathcal{G}_2$ and $\Gamma_3$ are non-local functions of density, with corresponding bias parameters. We neglect here a higher derivative term proportional to $\nabla^2\delta_{cb}$ which is exactly degenerate with an EFTofLSS counter-term. The stochastic contribution is given here by $\epsilon$, which is taken to have a power spectrum of the form
\be
P_{\epsilon\epsilon}(\bfk) = N + e_0 k^2 + e_2 k^2\mu^2\,,
\ee
where we assume the constant part, $N$, to include the Poisson shot noise and possible deviations from it, and we add two scale-dependent terms with different redshift-space dependence. Here and elsewhere, $\mu\equiv \hat \bfk \cdot \hat \bfn$ is the cosine of the angle between the wave-vector $\bfk$ and the line-of-sight direction $\hat \bfn$.
As is clear from our bias modelling, we follow the standard approximation for massive neutrinos in which the galaxy field is assumed to depend only on the density of cold matter, excluding neutrinos. This approach is also used in Refs.~\cite{Chudaykin:2020, Chen:2022jzq, DAmico:2020kxu}, and is based on the simulation-based results of Refs.~\cite{Castorina:2015bma,Bayer:2021kwg}.

We use the EFTofLSS to include the effects of small scales on mildly nonlinear ones by adding counter-terms to the redshift-space power spectrum, which is based on 1-loop perturbation theory~\cite{Bernardeau:2001qr}. This results in the following expression for the galaxy power spectrum:
\begin{align}
\label{eq:eftii}
P_{gg}(\bfk) = &Z_1^2(\bfk) P_L(k) + 2 \int \de^3 \bfq \, \left[ Z_2(\bfq, \bfk - \bfq) \right]^2P_L(q) P_L(\vert \bfk - \bfq \vert) \nonumber \\
&+ 6 Z_1(\bfk)P_L(k) \int \de^3 \bfq \, Z_3(\bfk, \bfq, -\bfq)P_L(q) + P_{\rm ctr}(\bfk) +P_{\epsilon\epsilon}(\bfk)\,,
\end{align}
where the redshift-space kernels $Z_1(\bfk)$, $Z_2(\bfk_1, \bfk_2)$, $Z_3(\bfk_1, \bfk_2, \bfk_3)$ can be found in Eq.~(2.14) of Ref.~\cite{Ivanov:2019pdj}, which includes more details and a derivation with the same bias basis used here (see also Ref.~\cite{Scoccimarro:1999}). We denote the linear power spectrum of $\delta_{cb}$ as $P_L(k)$ and have omitted the redshift dependence for brevity. We use the EdS approximation when computing these kernels, such that their redshift dependence is encoded in the linear growth factor, $D$, and the linear growth rate, $f$. This approximation has been shown to be accurate at better than 1\% for $\Lambda$CDM at the redshifts of BOSS (see e.g. Ref.~\cite{Donath:2020abv} for a detailed analysis). The contribution from counter-terms is given by
\begin{align}
P_{\rm ctr}(k,\mu) = &- 2 \tilde{c}_0 k^2 P_L(k) - 2 \tilde{c}_2 k^2 f \mu^2 P_L(k) - 2 \tilde{c}_4 k^2 f^2 \mu^4 P_L(k) \nonumber\\
&+c_{\nabla^4 \delta} k^4 f^4 \mu^4 (b_1 + f \mu^2)^2 P_L(k)\,,
\end{align}
where we define four counter-term parameters, resulting in eleven nuisance parameters in total.

Infrared re-summation is performed by splitting the linear power spectrum in a smooth, broadband component $P^{\rm nw}_L$ and a wiggle component $P^{\rm w}_L$. This wiggle-no-wiggle decomposition is performed using an Eisenstein-Hu fitting function~\cite{Eisenstein:1997ik}, denoted by $P_{\rm EH}$, to which a Gaussian filter is applied to improve the agreement with the linear spectrum, $P_L$. This results in a no-wiggle spectrum given by
\be
P^{\rm nw}_L(k)= P_{\rm EH}(k)\int_{\log_{10} k-4\lambda}^{\log_{10} k+4\lambda}{{\rm d}\log_{10} q}\frac{1}{\sqrt{2\pi\lambda^2}}\exp{\left[-\frac{\left(\log_{10} (q/k)\right)^2}{2\lambda^2}\right]}\frac{P_L(q)}{P_{\rm EH}(q)}\,,
\ee
where $\lambda$ is the width of the filter, which we choose to be $\lambda=0.25$. We then evaluate the loop terms in Eq.~\eqref{eq:eftii} with both the usual linear power spectra and $P^{\rm nw}_L$, and construct a wiggle version of the loop terms by subtracting the two. This, together with the linear contribution $P_{\rm w}$, constitutes $P_{gg}^{\rm w}$. We then apply the appropriate damping factors to the 1-loop and linear terms and sum the results via
\be
P_{gg}^{\rm re-sum}=P_{gg}^{\rm nw} + e^{-k^2\Sigma_{\rm tot}^2}  P_{gg}^{\rm w} + e^{-k^2 \Sigma_{\rm tot}^2}k^2 \Sigma_{\rm tot}^2\left(b_1+f\mu^2\right)^2 P_L^{\rm w}(k)\,, \label{eq:resum}
\ee
where  $\Sigma_{\rm tot}$ is given by Eq. (2.30) of Ref.~\cite{Chudaykin:2020}, which is itself based on the work of Ref.~\cite{Blas:2016sfa}.

Due to the assumption of a particular fiducial cosmology in the measurement of the data, we apply the Alcock-Paczynski (AP) rescaling to our theoretical prediction. This is done in the standard way, by modifying the scales and angles via
\begin{align}
    \bar{k}^2=k^2\left(\frac{H_0^{\rm fid}}{H_{0}}\right)^2\left(\left(\frac{H}{H^{\rm fid}}\right)^2 \mu^2+\left(\frac{D_{A}^{\rm fid}}{D_A}\right)^2(1-\mu^2)\right)\,,\\
    \bar{\mu}^2=\mu^2\left(\frac{H}{H^{\rm fid}}\right)^2\left(\left(\frac{H}{H^{\rm fid}}\right)^2 \mu^2+\left(\frac{D_{A}^{\rm fid}}{D_A}\right)^2(1-\mu^2)\right)^{-1}\,,
\end{align}
and applying a rescaling to the power spectrum of
\be
A_{\rm AP}=\left(\frac{H_0^{\rm fid}}{H_{0}}\right)^3\frac{H}{H^{\rm fid}}\left(\frac{D_{A}^{\rm fid}}{D_A}\right)^2\,,
\ee
where all quantities are evaluated at the effective redshift of the sample, and fiducial quantities are evaluated in the fiducial cosmology, which is a flat $\Lambda$CDM cosmology with $\Omega_m=0.31$, $H_0=67.6$, $\Omega_b h^2=0.022$ and $M_\nu=0.06~{\rm eV}$~\cite{Beutler_2016}.

After the application of the AP distortion we compute the multipoles of the power spectrum with:
\be
P_\ell(k)=\frac{2\ell+1}{2}\int_{-1}^1{\rm d} \mu\, P\left(\bar{k}(k,\mu),\bar{\mu}(k,\mu)\right)\mathcal{P}_\ell(\mu)\,,
\ee
where $\mathcal{P}_\ell(\mu)$ is the Legendre polynomial of order $\ell$. We follow Ref.~\cite{Chudaykin:2020} and re-define three of our counter-term parameters to be the combinations arising for each multipole, resulting in 
\be
c_0=\tilde{c}_0+\frac f 3 \tilde{c}_2 +\frac {f^2}{5} \tilde{c}_4\,,\ c_2=\tilde{c}_2+\frac {6f} {7} \tilde{c}_4\,,\ c_4=\tilde{c}_4\,.
\ee
The full set of 11 nuisance parameters per redshift bin that we use is therefore
\be
\theta_{\rm nuis}=\{b_1,b_2,b_{\mathcal{G}_2},b_{\Gamma_3},N,e_0,e_2,c_0,c_2,c_4,c_{\nabla^4\delta}\}\,.
\ee

To accelerate our analysis, the linear power spectrum is computed using the \texttt{bacco} emulator~\cite{Arico:2021izc,Angulo:2020vky,Arico:2020lhq}. This provides accurate evaluations of the output of Boltzmann solvers for general cosmologies in a fraction of the time, giving order 100 speed-up of the linear calculations. The calculation of the growth is done via a fast numerical integration of the linear version of Eqs.~\eqref{dens_eq}, \eqref{vel_eq}. The integrals of the redshift-space loop corrections are computed using a custom implementation based on the {\sc Fast-PT} algorithm~\cite{McEwen:2016,Fang:2017}, which also improves the speed of our theoretical predictions substantially with respect to standard integration techniques. We note that this is an independent pipeline from those publicly available, such as \texttt{CLASS-PT}~\cite{Chudaykin:2020} and \texttt{pybird}~\cite{DAmico:2020kxu}. The python-based theory code employed here is part of a complete likelihood pipeline, which will be made available in a dedicated paper~\cite{PBJ_paper}. In this work, we use only the theoretical prediction of the power spectrum and a distinct likelihood constructed specifically to analyse this data. However, this code includes also a full likelihood for the joint analysis of the power spectrum and bispectrum, which has been used in Refs.~\cite{Rizzo:2022lmh,Tsedrik:2022cri}.

To predict the BAO signal, we compute the Hubble rate and angular diameter distance numerically with a simple code validated against CLASS~\cite{Blas_2011} and CAMB~\cite{Lewis:1999bs}, and obtain the size of the sound horizon at the drag epoch using a fitting formula from Ref.~\cite{Aubourg:2014yra}, given by
\be
r_d=\frac{55.154 \exp\left[-72.3 (\omega_\nu+0.0006)^2\right]}{\omega_b^{0.12807} (\omega_c+\omega_b)^{0.25351}}\,,
\ee
where we also approximate the neutrino density via $\omega_\nu=0.0107~M_\nu/{\rm eV}$. We then compute the different combinations of distances as required by the different datasets available. 

\subsubsection{Analysis set-up}

We start by considering a Gaussian likelihood constructed with the power spectrum multipoles, the BAO signal and including their cross covariance where relevant. We then perform analytical marginalization over the 8 nuisance parameters that appear linearly in the power spectrum, which are $b_{\Gamma3},N,e_0,e_2,c_0,c_2,c_4,c_{\nabla^4\delta}$. We follow the procedure explained in Refs.~\cite{DAmico:2019fhj,DAmico:2020kxu} and thus obtain a modified posterior distribution which is no longer Gaussian with respect to the data vector, but drastically reduces the dimension of the sampled parameter space, accelerating the convergence of the MCMC analysis. 

In our main analyses, we include multipole data from all 3 multipoles from 2 redshifts and 2 sky cuts up to a maximum wavenumber of $k_{\rm max}=0.2~ h/{\rm Mpc}$ as well as all BAO data, including that from the same redshifts and sky cuts as the multipoles and the data from lower ($z=0.106,0.15$) and higher redshifts ($z=2.334$). We also perform simplified analyses without BAO data to evaluate the impact of that dataset on the constraints. 

We run 3 different sets of analysis, one for each model of interest: $\Lambda$CDM, $w$CDM, and  $wA$CDM. We vary 5 standard cosmological parameters, $\log(10^{10} A_s)$, $n_s$, $\omega_c$, $\omega_b$ and $h$ in all analyses, as well as the dark energy EOS parameter, $w$, and the interaction strength $A$ as defined in Eq.\eqref{A_def}, in the relevant cases. Throughout our work, we adopt a BBN prior on $\omega_b$. Some of our main analyses include a $3\sigma$ Planck prior on the early Universe parameters $\log(10^{10}A_s)$ and $n_s$, but we also run CMB-free analyses by relaxing this assumption. Furthermore we fix the sum of the neutrino masses to $M_\nu=0.06~{\rm eV}$ and use broad flat priors for the remaining cosmological parameters. Specifically, for $\omega_c$ and $h$ we impose flat priors matching the ranges of parameters allowed by the \texttt{bacco} linear emulator, which are
\be
\Omega_{cb}\sim \mathcal{U}(0.06,0.7)\,,\ h\sim \mathcal{U}(0.5,0.9)\,,
\ee
in which $\mathcal{U}(a,b)$ denotes a uniform distribution with edges $a,b$ and $\Omega_{cb}=\Omega_c+\Omega_b$. For numerical reasons, we also impose flat priors on $w$ and $A$, given by
\be
w\sim \mathcal{U}(-3,-0.5)\,,\ \frac{A}{{\rm [b/GeV]}}\sim \mathcal{U}(-10^3,10^3)\,.
\ee
Finally, these parameters also have to obey the physical condition that the dark scattering cross-section is positive, giving the prior $\xi\sim \mathcal{U}(0,\infty)$. This translates into a prior on $w$ and $A$ that forces ${\rm sign}(A)={\rm sign}(1+w)$, which gives rise to a characteristic butterfly shape of the posteriors, as seen in Refs.~\cite{Carrilho:2021hly,Tsedrik:2022cri}.

Our baseline priors for the nuisance parameters are the same as those in Ref.~\cite{Philcox:2021main} and are as follows:
\begin{gather}
    b_1\sim \mathcal{U}(0,4)\,,\ b_2\sim\mathcal{N}(0,1)\,,\ b_{\mathcal{G}_2}\sim\mathcal{N}(0,1)\,,\ b_{\Gamma_3}\sim\mathcal{N}\left(\frac{23}{42}(b_1-1),1\right)\,,\nonumber\\
    N\sim\mathcal{N}\left(\frac{1}{\bar{n}},\frac{2}{\bar{n}}\right)\,,\  e_0\sim\mathcal{N}\left(0,\frac{2}{\bar{n}k_{\rm NL}^2}\right)\,,\ e_2\sim\mathcal{N}\left(0,\frac{2}{\bar{n}k_{\rm NL}^2}\right)\,,\
    \frac{c_0}{[{\rm Mpc}/h]^2}\sim\mathcal{N}(0,30)\,,\label{eq:prior_base}\\ \frac{c_2}{[{\rm Mpc}/h]^2}\sim\mathcal{N}(30,30)\,,\ \frac{c_4}{[{\rm Mpc}/h]^2}\sim\mathcal{N}(0,30)\,,\ 
    \frac{c_{\n^4\delta}}{[{\rm Mpc}/h]^4}\sim\mathcal{N}(500,500)\,,\nonumber
\end{gather}
in which $\mathcal{N}(\mu,\sigma)$ denotes a normal distribution with mean $\mu$ and standard deviation $\sigma$. We use a nonlinear scale $k_{\rm NL}=0.45~h/{\rm Mpc}$ and we assume an inverse number density of $\bar n^{-1}=3500~({\rm Mpc}/h)^3$ for the $z_1=0.38$ bin and $\bar n^{-1}=5000~({\rm Mpc}/h)^3$ for the $z_3=0.61$ bin. 

While many of these parameters are marginalised over analytically and do not need to be sampled over in the analysis, it is important to set priors on them since they multiply nonlinear contributions, which should not be larger than the linear contributions which are well known to be accurate on sufficiently large scales.\footnote{The choice of Gaussian priors instead of flat ones is exactly to allow for analytical marginalisation, as that requires Gaussian integrals to be performed.} These particular choices of priors are informed by the arguments of Refs.~\cite{Chudaykin:2020, Ivanov:2021kcd}, which are based on the validity of the perturbative EFT modelling used here. In particular, bias parameters are expected to be $O(1)$ and noise parameters are expected to be the same order as the inverse number density and to scale with the appropriate power of the nonlinear scale. The counter-terms are also expected to be $O(1)$ in units of $k_{\rm NL}$, but it is argued in Ref.~\cite{Chudaykin:2020} that for this effect the relevant nonlinear scale is larger, given the rather large velocity dispersion values ($\sigma_v\sim 5~{\rm Mpc}/h$) found in previous BOSS analyses~\cite{Beutler_2016}. This justifies using standard deviations of $30~({\rm Mpc}/h)^2$ and $500~({\rm Mpc}/h)^4$ for the quadratic and quartic counter-terms respectively. While some of these expectations for bias parameters are supported by measurements from simulations~\cite{Lazeyras:2015lgp,Rizzo:2022lmh,Zennaro:2021pbe,Barreira:2021ukk,Tsedrik:2022cri}, this is less so for the counter-term parameters, for which it is much more difficult to get definite predictions. These are our baseline priors used in most of our analyses. However, we also perform additional analyses in Appendix~\ref{sec:prior} to test the relevance of these assumptions by enlarging the variance of these priors by factors 3 and 10. This will also allow us to include potential caveats to the arguments presented above due to the modified cosmologies which we study here. As is shown there, at least for the BOSS data, these priors appear to be informative and a further study of their effects is essential for obtaining robust results.

We sample our posterior via an MCMC analysis using
Goodman \& Weare’s affine invariant sampler as implemented in \texttt{emcee}~\cite{Foreman_Mackey_2013}. We use \texttt{GetDist}~\cite{Lewis:2019xzd} to plot our posterior distributions. In each case, we compute the best-fit parameters by maximizing the analytically marginalised posterior.

\section{Results}
\label{sec:resultsmain}

We now present our results for our main analyses for the three cosmological models under study. We begin with $\Lambda$CDM, demonstrating that we reproduce the results of Ref.~\cite{Philcox:2021main}. Next we add dynamical dark energy with a constant $w$ and finally we show our results for the dark scattering model, which we denote $wA$CDM, where we show the first ever constraints on the interaction strength $A$.

\subsection{$\Lambda$CDM analysis}

\begin{figure}[tbp]
\centering 
\includegraphics[width=\textwidth]{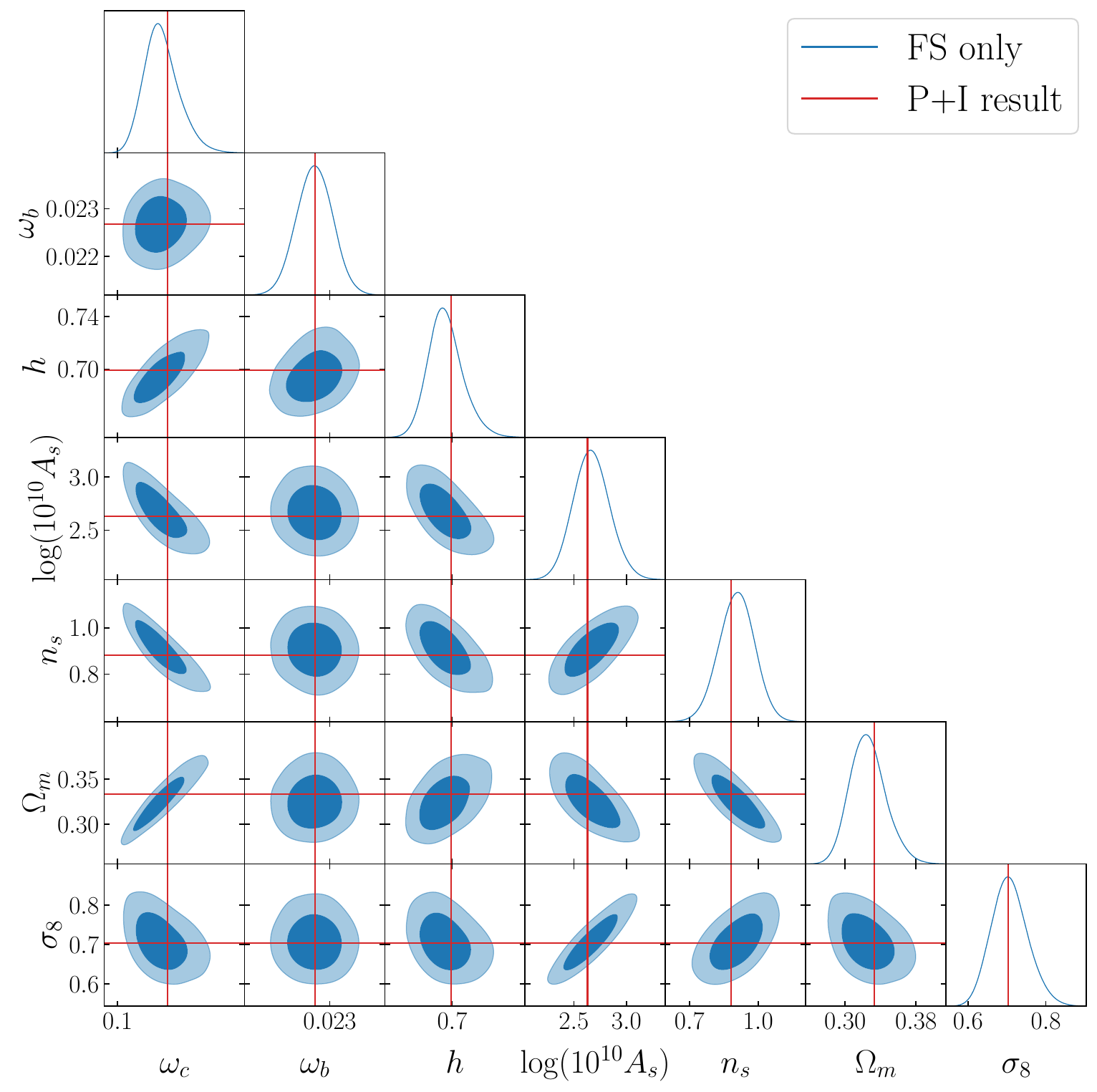}
\caption{\label{fig:lcdm_comp_PI} Marginalized posteriors for the sampled cosmological parameters and $\Omega_m$ and $\sigma_8$ for $\Lambda$CDM for the analysis using only the full-shape power spectrum data. We use $k_{\rm max}=0.20~h/{\rm Mpc}$ for all 3 multipoles. This case mimics the analysis of Philcox and Ivanov (P+I)~\cite{Philcox:2021main}, and solid red lines show the mean of the parameters obtained there, shown in Table III of that reference, demonstrating perfect agreement.}
\end{figure}

We present the results of our full-shape analysis for $\Lambda$CDM in Fig.~\ref{fig:lcdm_comp_PI}, showing the marginalised posteriors we obtain, compared to the means for the results of Ref.~\cite{Philcox:2021main} (denoted by P+I), which performed the first analysis with the data measured with the estimator used here. We also show a summary of those results in Table~\ref{tab_lcdm_comp_PI}. It is clear that our results are in perfect agreement with those of Ref.~\cite{Philcox:2021main}, using an entirely independent likelihood pipeline and theory code, with only the EFT and bias modelling being the same. We consider this as a validation test of our theory and likelihood code and its application to data. We will now show and discuss our main results.

\begin{table}[h]
\centering
 \begin{tabular}{||l l l ||} 
 \hline 
 Parameter & FS only [this work] & P+I analysis~\cite{Philcox:2021main} \\
 \hline\hline
$\omega_c$ & $0.134^{+0.011}_{-0.015}$ & $0.139^{+0.011}_{-0.015}$ \\  
 \hline
 $h$ & $0.695^{+0.011}_{-0.015} $ & $0.699^{+0.015}_{-0.017}$ \\ 
 \hline
$\log(10^{10} A_s)$ & $2.67\pm 0.17$  & $2.63^{+0.15}_{-0.16}$ \\  
 \hline
 $n_s$ & $0.906\pm 0.078 $  & $0.883^{+0.076}_{-0.072}$\\ 
 \hline\hline
 $\Omega_m$ &  $0.326^{+0.017}_{-0.022}$ & $0.333^{+0.019}_{-0.020}$ \\ 
 \hline
 $\sigma_8$ &  $0.707^{+0.042}_{-0.051}$  & $0.704^{+0.044}_{-0.049} $ \\
 \hline
\end{tabular}
\caption{Parameter constraints for the $\Lambda$CDM analysis using full shape data only compared to the results of Ref.~\cite{Philcox:2021main} for the equivalent analysis. We show the 5 sampled parameters as well as 2 derived parameters, $\Omega_m$, and $\sigma_8$. The values shown are the means with the $1\sigma$ confidence intervals.}
\label{tab_lcdm_comp_PI}
\end{table}

\begin{figure}[tbp]
\centering 
\includegraphics[width=\textwidth]{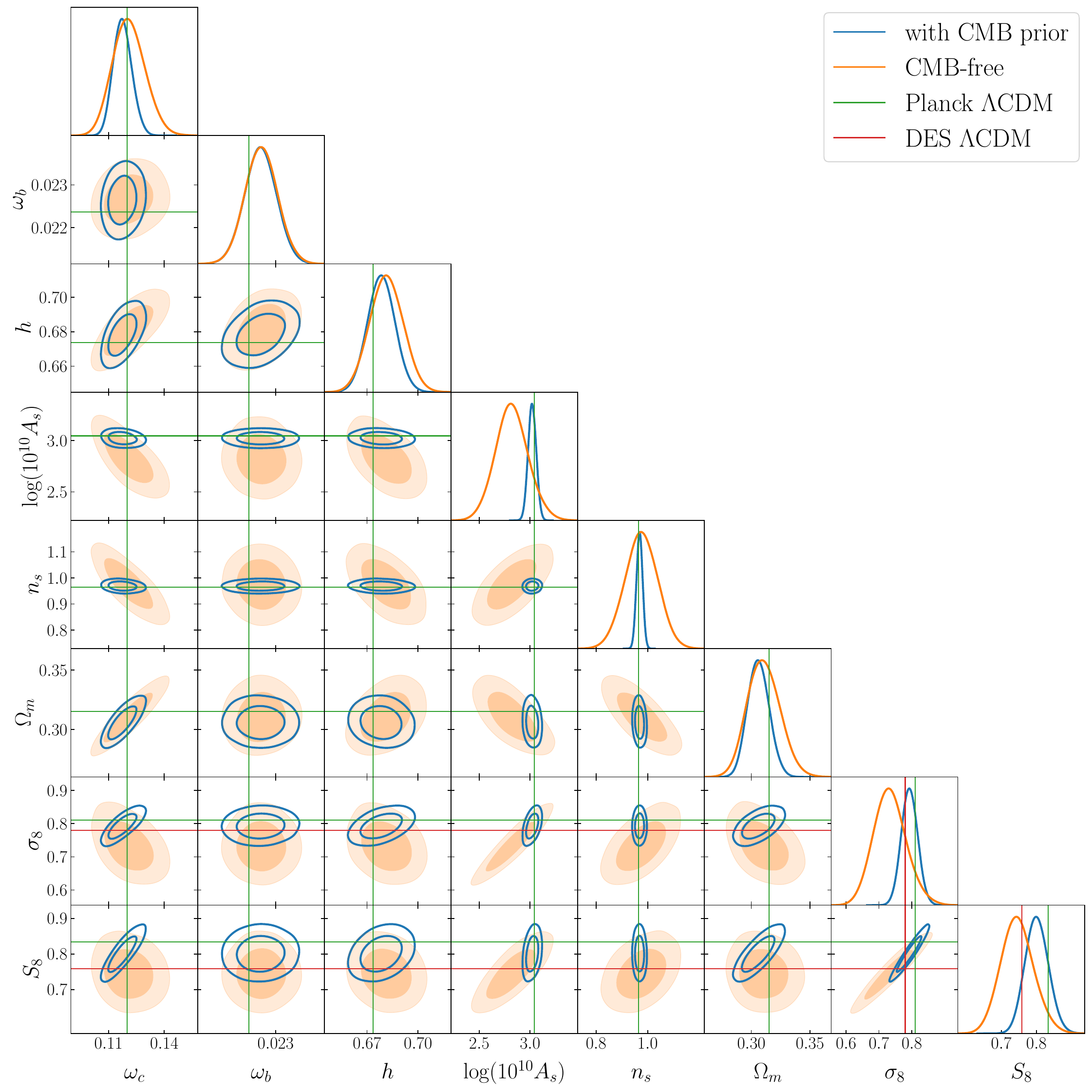}
\caption{\label{fig:lcdm} Marginalized posteriors for the sampled cosmological parameters for $\Lambda$CDM, showing both the analysis with (blue empty contours) and without (orange filled contours) a 3$\sigma$ Planck prior on the primordial parameters (as well as a BBN prior on $\omega_b$). We use $k_{\rm max}=0.20~h/{\rm Mpc}$ for all 3 power spectrum multipoles and use the full-shape data and the BAO measurement. Solid green lines mark the best-fit values from the Planck $\Lambda$CDM analysis, while red lines show the $\sigma_8$ and $S_8$ measured in the fiducial $\Lambda$CDM DES analysis~\cite{DES:2021bvc, Secco:2021vhm}.}
\end{figure}

Our main $\Lambda$CDM analyses are shown in Fig.~\ref{fig:lcdm}, with the summary results reported in Table~\ref{tab_baseline_lcdm}. There we show both the case with a CMB prior on the primordial parameters $A_s$ and $n_s$, as well as the CMB-free analysis, similar to the one shown in Fig.~\ref{fig:lcdm_comp_PI}. Both these analyses differ from the previous one in the inclusion of the measurement of the BAO scale at various redshifts ranging from 0.15 to 2.334. The addition of this information constrains the expansion history better, and therefore provides a sharper measurement of $\omega_c$ and $h$. For this reason, the degeneracy between $\omega_c$ and $n_s$ is somewhat broken, thus moving the measurement of $n_s$ to larger values as $\omega_c$ is lowered, placing it in agreement with the Planck measurement, even in the CMB-free analysis.

\begin{table}[h]
\centering
 \begin{tabular}{||l l l ||} 
 \hline 
 Parameter & $\Lambda$CDM CMB prior & $\Lambda$CDM CMB-free \\
 \hline\hline
$\omega_c$ & $0.118\pm0.005$ ($0.120$) & $0.121^{+0.008}_{-0.009} $ ($0.120$) \\  
 \hline
 $100 \omega_b$ & $2.264\pm 0.037$ ($2.256$) & $2.266\pm 0.038 $ ($2.244$)\\ 
 \hline
 $h$ & $0.678\pm 0.008 $ ($0.681$) & $0.681\pm 0.010$ ($0.678$)\\ 
 \hline
$\log(10^{10} A_s)$ & $3.021\pm 0.040 $ ($3.023$)  & $2.821\pm 0.158 $ ($2.832$)\\  
 \hline
 $n_s$ & $0.969\pm 0.012 $ ($0.974$) & $0.975\pm 0.063$ ($0.972$) \\ 
 \hline\hline
 $\Omega_m$ &  $0.306\pm 0.009$ ($0.308$) & $0.311^{+0.013}_{-0.015}$ ($0.311$)\\ 
 \hline
 $\sigma_8$ &  $0.793\pm 0.025$ ($0.806$) & $0.734^{+0.046}_{-0.053}$ ($0.731$)\\ 
 \hline
 $S_8$ &  $0.801\pm 0.033$ ($0.816$) & $0.746^{+0.044}_{-0.049}$ ($0.744$) \\ 
 \hline
\end{tabular}
\caption{Parameter constraints for the two $\Lambda$CDM analyses using full shape data as well as measurements of the BAO scale (FS+BAO). We show the 5 sampled parameters as well as 3 derived parameters, $\Omega_m$, $\sigma_8$ and $S_8$. The values shown are the means with the $1\sigma$ confidence intervals and the best-fit value in parenthesis.}
\label{tab_baseline_lcdm}
\end{table}

The two analyses are broadly consistent, except for the measured value of the scalar amplitude, which is substantially lower in the CMB-free case, with a correspondingly low value of $\sigma_8$ being measured. The value measured in the full-shape + BAO analysis (FS+BAO) is approximately $1\sigma$ larger than the value measured in the FS only analysis discussed above ($\log (10^{10} A_s) = 2.67$ for FS vs $2.82$ for FS+BAO). This is a consequence of the more precise measurement of $\omega_c$ whose degeneracy with $A_s$ raises the value of the amplitude. In spite of this, the scalar amplitude is still over $1\sigma$ away from the value measured by Planck, with $\sigma_8$ agreeing more closely with the values measured by weak lensing experiments such as DES~\cite{DES:2021bvc, Secco:2021vhm}. 

Low values of the scalar amplitude have been found in various previous full-shape analyses of the BOSS DR12 data~\cite{Philcox:2021main,Troster:2019ean,Chudaykin:2020ghx,DAmico:2020kxu,Ivanov:2019pdj,Chen:2022jzq,DAmico:2019fhj,Chen:2021wdi}. Precise comparisons are complicated by different choices of data, estimators and priors used, as well as the fact that some of these analyses are likely to suffer from the window function normalisation issue discussed in Refs.~\cite{deMattia:2019vdg,deMattia:2020fkb,Beutler:2021eqq} and therefore may find a smaller amplitude than Planck for that reason. Our FS+BAO analysis is more similar to that conducted by Ref.~\cite{Chudaykin:2020ghx}, which also use BAO data in a broad redshift range and find $\Lambda$CDM results consistent with ours. Ref.~\cite{DAmico:2020kxu} also uses similar BAO data but focuses only on beyond-$\Lambda$CDM analyses, which we discuss below in detail. Other analyses which set a Planck prior on $n_s$~\cite{Chen:2021wdi,Philcox:2021main} but do not use the same BAO data also obtain very similar results for the amplitude as our FS+BAO CMB-free analysis of $\sigma_8=0.733 \pm 0.047$ and $\sigma_8=0.729^{+0.036}_{-0.042}$. This similarity is likely because adding the BAO data effectively moves the values of $\omega_c$ and consequently of $n_s$ close to the Planck value and thus results in consistent values for all parameters. It is worth noting that in both of these works, the normalisation of the window function is correctly accounted for.

It is important also to compare our results with those of the official BOSS and eBOSS collaborations as well as those not using EFT-based modelling. The latest official combined results of Ref.~\cite{Alam:2020sor} show a value of the late-time amplitude of $\sigma_8=0.838 \pm 0.059~(8.48)$ based on RSD measurements alone, which is in agreement with the Planck result and is therefore in some disagreement with our results. Those results combine BOSS galaxies and eBOSS galaxies and quasars and use template fitting instead of the full-shape analysis done here, as well as modelling nonlinear RSDs via the TNS model~\cite{Taruya:2010mx}.

A similar BOSS + eBOSS analysis using full-shape information was performed in Ref.~\cite{Semenaite:2021aen}, which finds $\sigma_8=0.815\pm0.044$ (or $\sigma_8= 0.800\pm0.039$ with tight priors on $\omega_b$, $\omega_c$ and $n_s$) for the full data combination. There, an analysis is also reported using only the BOSS DR12 data, finding lower values of $\sigma_8$ at $1.1\sigma$ tension with Planck and in broad agreement with our result. A similar DR12-only analysis had also been performed previously by Ref.~\cite{Troster:2019ean}, finding an even lower value of the amplitude of $\sigma_8=0.710\pm0.049$, also in broad agreement with our result. Both of these analyses are based on renormalised perturbation theory and a version of the TNS model, differing in the fact that Ref.~\cite{Semenaite:2021aen} uses a full co-evolution relation in their bias model whereas Ref.~\cite{Troster:2019ean} uses only a partial one, and likely also in the window normalisation.

Another recent analysis of BOSS + eBOSS was performed by Ref.~\cite{Neveux:2022tuk}, where the TNS model is also used, but an iterative emulator for the likelihood is employed. They obtain a joint constraint of $\sigma_8=0.877 \pm 0.051$ (with a prior on $\omega_b$ only), which is $1.5\sigma$ discrepant with Planck, but in the opposite direction of our results. They also show that this is driven by the quasar sample, which prefers a much higher value of $\sigma_8$, while the BOSS + eBOSS galaxy samples instead prefer a value of $\sigma_8 = 0.738^{+0.056}_{-0.031}$ in good agreement with our results.

Given the results of Refs.~\cite{Semenaite:2021aen, Neveux:2022tuk}, it is possible that our discrepancy with the official analysis of Ref.~\cite{Alam:2020sor} is due to the preference of the higher redshift sample for a larger amplitude and/or due to differences between full-shape and template-based analyses. This difference between types of analyses has been explored using the recent \textit{ShapeFit} technique~\cite{Brieden:2021edu,Brieden:2021cfg,Brieden:2022lsd}. These works have also found that the quasar sample prefers a larger amplitude than the galaxy samples, in slight tension with Planck. However, when their analyses exclude quasars, they find for eBOSS LRGs $\sigma_8=0.820 \pm 0.043$~\cite{Brieden:2022lsd} and for BOSS DR12, $\sigma_8=0.806 \pm 0.065$~\cite{Brieden:2021cfg}. The latter analysis is comparable to ours, shows a $1.1\sigma$ deviation from our result and is in agreement with Planck. Beyond this, Fig.~1 of Ref.~\cite{Brieden:2021cfg} directly compares an EFT-based full-shape method with its \textit{ShapeFit} technique and shows that the full-shape results give a lower value of $\sigma_8$, which is likely in agreement with our results shown here. This is another indication that full-shape analyses prefer somewhat lower amplitude values, at least at the level of BOSS DR12 data.

Finally, we point out Ref.~\cite{DAmico:2022osl}, where an EFT-based full-shape analysis of BOSS DR12 is performed that does not find a low amplitude. That work, however, argues that projection effects are the cause of the low amplitudes found in other analyses and corrects them using simulation-based shifts of parameters as well as modifying the priors of cosmological parameters. It is therefore not directly comparable to our results. In spite of this, it does raise the question of whether full-shape or EFT-based analyses are more susceptible to these projection effects, which may explain why template-based analyses find larger values of the amplitude, something which we aim to explore in future work.\\

Our analysis with a Planck prior on $A_s$ and $n_s$, essentially approximates the inclusion of CMB data and therefore results in all parameters being in agreement with Planck constraints. However, because a lower scalar amplitude is preferred by our analysis of the BOSS data, the best-fit value of $\sigma_8$ is still lower than in Planck, as is the value of the matter density. This preference is shown even in $A_s$ which, while still dominated by its prior ($\log(10^{10}A_s)=3.044\pm0.042$) is $0.5\sigma$ lower with a best-fit value of $\log(10^{10}A_s)=3.023$. 

In Appendix~\ref{sec:prior}, we discuss how these results depend on the choice of priors for the EFTofLSS nuisance parameters. In the CMB-free analysis, shown in Fig.~\ref{fig:cmb_free_prior_lcdm}, we find that a large broadening of the prior ranges ($10\times\sigma$) can change the results substantially, particularly for $n_s$, but also for the amplitude $A_s$. This is driven by the fact that the nuisance parameters are drawn towards values considered extreme by most current literature. An increase of the prior by a factor of 3 has a smaller impact, but does change the primordial parameters somewhat. For the CMB-dependent cases shown in Fig.~\ref{fig:cmb_prior_prior_lcdm}, shifts in parameters are largest in $\omega_c$ and $h$, and can be significant, changing the results by $O(1)$~$\sigma$. And while all this is fueled by rather large values of nonlinear bias parameters, this indicates that the priors are informative and without knowledge of them, the inferred values of parameters such as the scalar amplitude or the matter density are highly dependent on the choice of prior. This prior dependence is likely to be related, at least partially, to the projection effects mentioned above and may also explain to some extent the differences between full-shape and template-based analyses. This and related investigations will be the subject of future work.

We show the full contours including bias parameters in Fig.~\ref{fig:LCDM_all_pars} of Appendix~\ref{sec:full_cont}, with summary statistic shown in Table~\ref{tab_all_lcdm}, where more details on our main analyses can be seen.

\subsection{$w$CDM analysis}

\begin{figure}[tbp]
\centering 
\includegraphics[width=\textwidth]{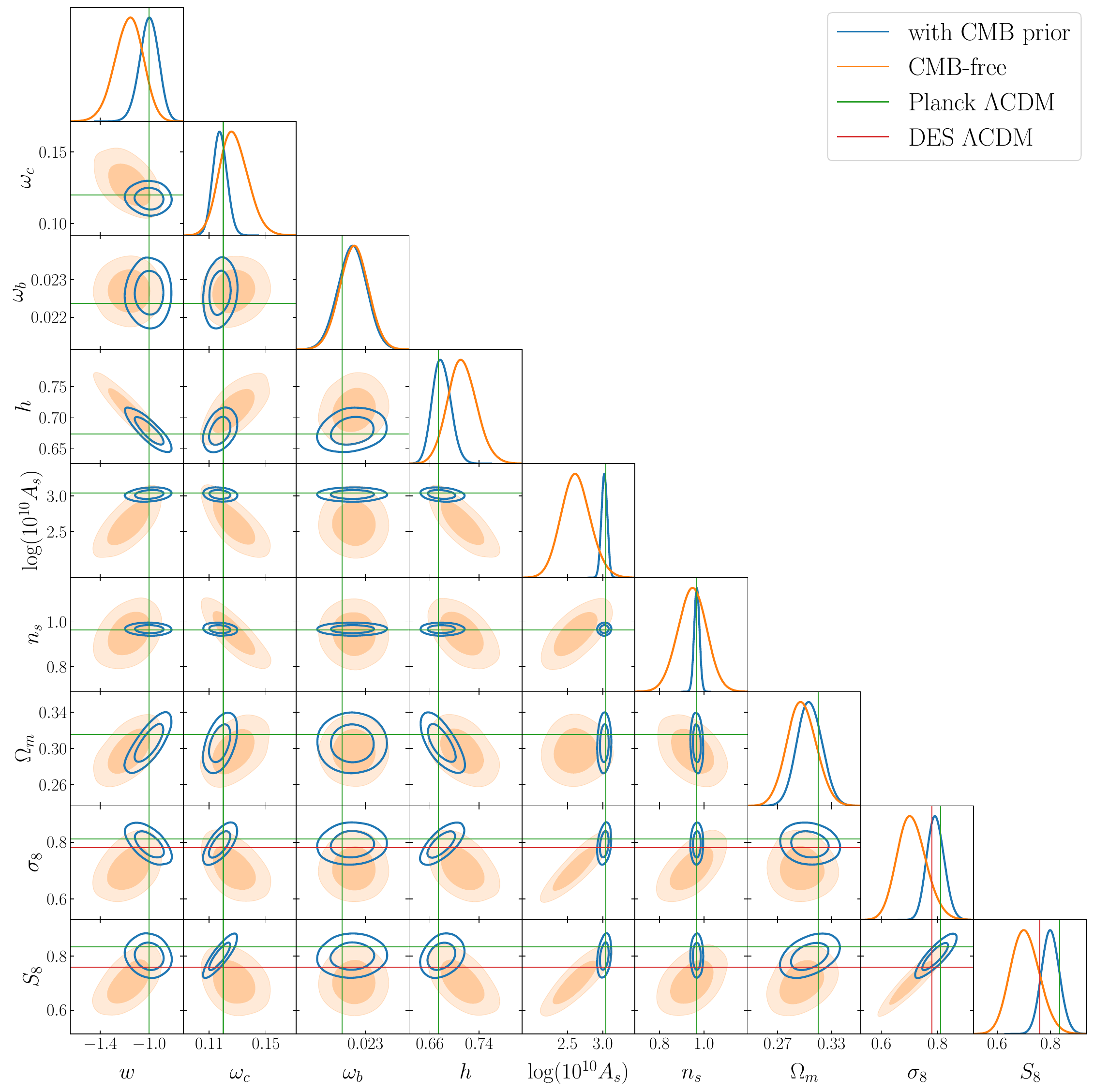}
\caption{\label{fig:wcdm} Marginalized posteriors for the sampled cosmological parameters for $w$CDM, showing both the analysis with (blue empty contours) and without (orange filled contours) a 3$\sigma$ Planck prior on the primordial parameters (as well as a BBN prior on $\omega_b$). We use $k_{\rm max}=0.20~h/{\rm Mpc}$ for all 3 multipoles and use the full shape data and the BAO measurement. Solid green lines mark the best-fit values from the Planck $\Lambda$CDM analysis while red lines show the $\sigma_8$ and $S_8$ measured in the fiducial $\Lambda$CDM DES analysis~\cite{DES:2021bvc, Secco:2021vhm}.}
\end{figure}

We now turn to the $w$CDM analysis, whose main results are shown in Fig.~\ref{fig:wcdm} and in Table~\ref{tab_baseline_wcdm}, and are obtained using both the full shape data and the BAO information. We show there both the CMB-free case and the one with a Planck prior on the primordial parameters. This Planck prior has a considerably larger impact here than in $\Lambda$CDM with many parameters having different posteriors between the two cases. This is due to the fact that in the CMB-free analysis a number of degeneracies are introduced when $w$ is varied, particularly with $h$ and $A_s$ but also with $\omega_c$. The BAO data plays a crucial role in reducing these degeneracies and without it, the background parameters would have very large error bars~\cite{DAmico:2020kxu,Chudaykin:2020ghx}. This can be seen below in our study of $wA$CDM, where we also show results without BAO.

The parameters measured in the CMB-free analysis show some deviation from the Planck results, most of which is driven by the low amplitude preferred by the BOSS data. The degeneracies mentioned above influence this too. We therefore find a rather low amplitude with $\log(10^{10} A_s)=2.62\pm 0.21$ and $\sigma_8 = 0.709^{+0.046}_{-0.055}$ as well as a preference for phantom dark energy, with $w=-1.17^{+0.12}_{-0.11}$. However, projection effects are exacerbating this deviation, given that the best-fit values are less discrepant than their means, resulting instead in $\log(10^{10} A_s)=2.77$, $\sigma_8=0.737$ and $w=-1.093$. 

Comparison with previous works is not straightforward, since previous analyses with the EFTofLSS of Refs.~\cite{DAmico:2020kxu,Chudaykin:2020ghx, DAmico:2020tty} do not use the same estimator and it is unclear whether the the power spectrum normalization is consistently applied~\cite{deMattia:2019vdg,deMattia:2020fkb,Beutler:2021eqq,Philcox:2021main}. Still, our results are broadly consistent with those of Ref.~\cite{DAmico:2020kxu}, particularly in terms of the best-fit values; the results of Ref.~\cite{Chudaykin:2020ghx} are consistent at the $1\sigma$ level; and, while Ref~\cite{DAmico:2020tty} focuses on the case with $w>-1$, their analysis with free $w$ is also broadly consistent with ours.

\begin{table}[h]
\centering
 \begin{tabular}{||l l l ||} 
 \hline 
 Parameter & $w$CDM CMB prior & $w$CDM CMB-free \\
 \hline\hline
 $w$ & $-1.002^{+0.081}_{-0.073}$ ($-1.023$) & $-1.17^{+0.12}_{-0.11} $ ($-1.093$) \\  
 \hline
 $\omega_c$ & $0.118\pm 0.005 $ ($0.118$) & $0.127^{+0.009}_{-0.011} $ ($0.122$) \\  
 \hline
 $100 \omega_b$ & $2.265\pm 0.038$ ($2.286$) & $2.269\pm 0.038  $ ($2.273$)\\ 
 \hline
 $h$ & $0.679^{+0.014}_{-0.016}$ ($0.689$) & $0.713\pm 0.024$ ($0.699$)\\ 
 \hline
$\log(10^{10} A_s)$ & $3.021\pm 0.041  $ ($3.027$)  & $2.62\pm 0.21 $ ($2.77$)\\  
 \hline
 $n_s$ & $0.968\pm 0.012 $ ($0.966$) & $0.947\pm 0.064$ ($0.9721$) \\ 
 \hline\hline
 $\Omega_m$ &  $0.306\pm 0.014 $ ($0.298$) & $0.297\pm 0.016$ ($0.297$)\\ 
 \hline
 $\sigma_8$ &  $0.793\pm 0.030$ ($0.802$) & $0.709^{+0.046}_{-0.055}$ ($0.737$)\\ 
 \hline
 $S_8$ &  $0.800\pm 0.033$ ($0.799$) & $0.705^{+0.050}_{-0.057}$ ($0.733$) \\ 
 \hline
\end{tabular}
\caption{Parameter constraints for the two $w$CDM analyses using full shape data as well as measurements of the BAO scale (FS+BAO). We show the 6 sampled parameters as well as 3 derived parameters, $\Omega_m$, $\sigma_8$ and $S_8$. The values shown are the means with the $1\sigma$ confidence intervals and the best-fit value in parenthesis.}
\label{tab_baseline_wcdm}
\end{table}

The analysis with a CMB prior on $A_s$ and $n_s$ once again moves all parameters closer to the values measured by Planck, as expected. In addition, the measured value of $w$ is now in full agreement with $\Lambda$CDM, being $w=-1.002^{+0.081}_{-0.073}$, a result which also agrees with the CMB + (e)BOSS analyses of Refs.~\cite{BOSS:2016wmc,Alam:2020sor,Brieden:2022lsd}. Furthermore, the errors on all parameters are substantially reduced relative to the CMB-free analysis, since degeneracies are alleviated or broken. As we have already shown in the corresponding $\Lambda$CDM analysis, the scalar amplitude is still pushed down somewhat, resulting in a slightly lower value of $\sigma_8$ than inferred from Planck.

Full contours including bias parameters can be seen in Fig.~\ref{fig:wCDM_all_pars} of Appendix~\ref{sec:full_cont}, with summary statistics shown in Table~\ref{tab_all_wcdm}.

We repeat the exercise done for $\Lambda$CDM of extending the prior range of the nuisance parameters also for $w$CDM, showing the results in Figs.~\ref{fig:cmb_free_prior_wcdm} and \ref{fig:cmb_prior_prior_wcdm} and Table~\ref{tab_prior_wCDM_bias} of Appendix~\ref{sec:prior}. The conclusions are very similar to the case of $\Lambda$CDM, but we find in addition that broader priors have a small effect on the constraint on $w$. In the CMB-dependent cases, the uncertainties on $w$ grow only slightly and agree with the base case shown here. However, in the CMB-free case, there is a small shift to $w=-1.20^{+0.13}_{-0.12}$ when the priors are 3 times larger and a larger shift to $w=-1.11^{+0.19}_{-0.12}$ for 10 times broader prior ranges. While relatively small, these shifts imply that the choice of priors directly affects the results of the analysis and one should therefore be cautious in the interpretation of the constraints.

\subsection{$wA$CDM analysis}

\label{sec:results}
\begin{figure}[tbp]
\centering 
\includegraphics[width=.9\textwidth]{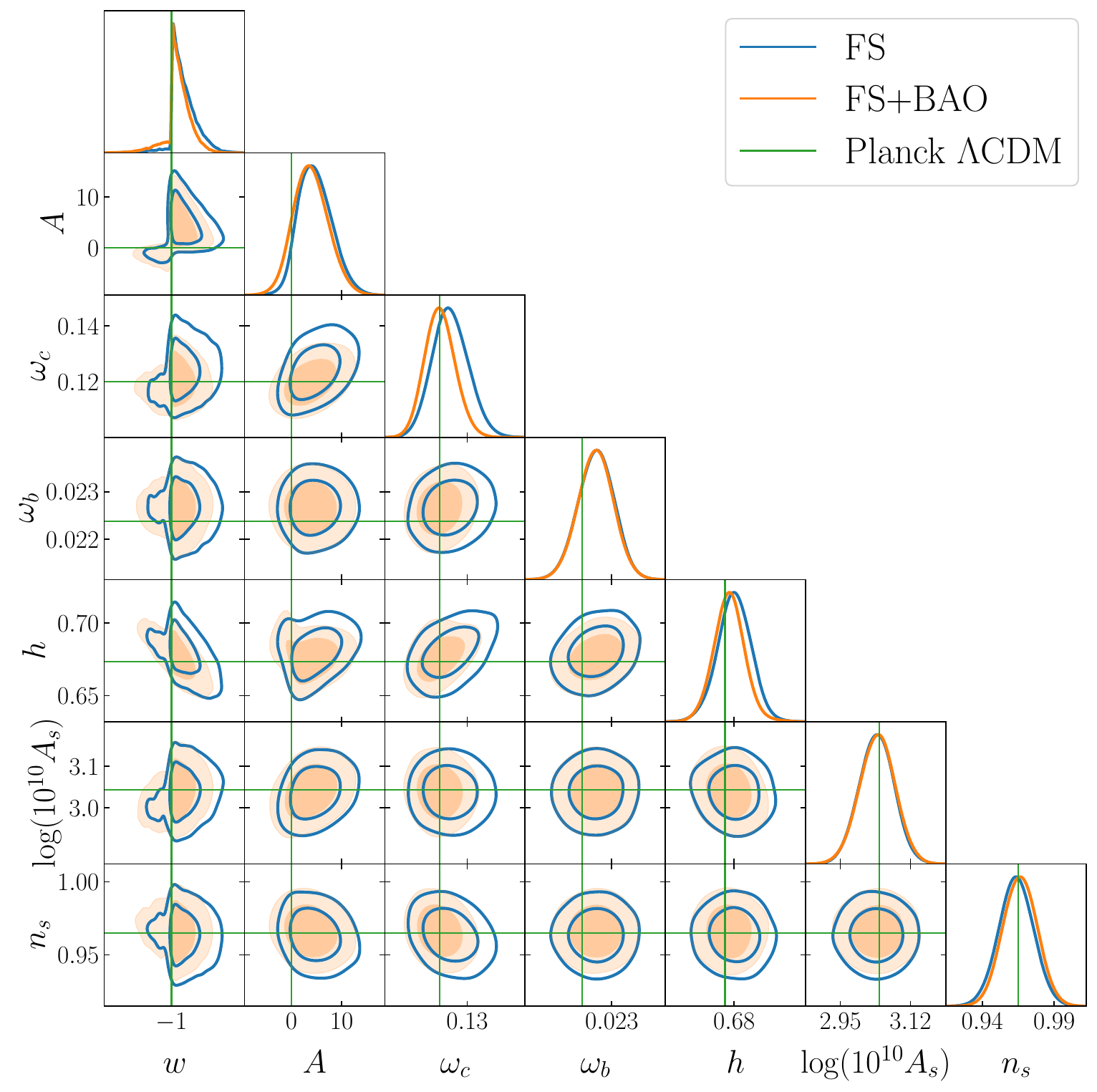}
\caption{\label{fig:baseline_BAO} Marginalized posteriors for the sampled cosmological parameters in the analysis with a Planck prior on the primordial parameters (as well as a BBN prior on $\omega_b$). We use $k_{\rm max}=0.20~h/{\rm Mpc}$ for all 3 multipoles and show both the analysis with the full shape only (FS, blue empty contours) and with both the full shape and the BAO measurement (FS+BAO, orange filled contours). The shape of the $w-A$ contour is due to the physical prior imposed on those parameters, i.e. ${\rm sign}(A)={\rm sign}(1+w)$. Solid green lines mark the best-fit values from Planck $\Lambda$CDM analysis.}
\end{figure}

Finally, we now describe our analyses for dark scattering, which we denote by $wA$CDM. We begin by presenting our baseline analysis, allowing all cosmological parameters to vary, but imposing a 3$\sigma$ Planck prior on $A_s$ and $n_s$; and leave the CMB-free analysis, which suffers from severe degeneracies, to the last part of this section.

\begin{figure}[tbp]
\centering 
\includegraphics[width=.9\textwidth]{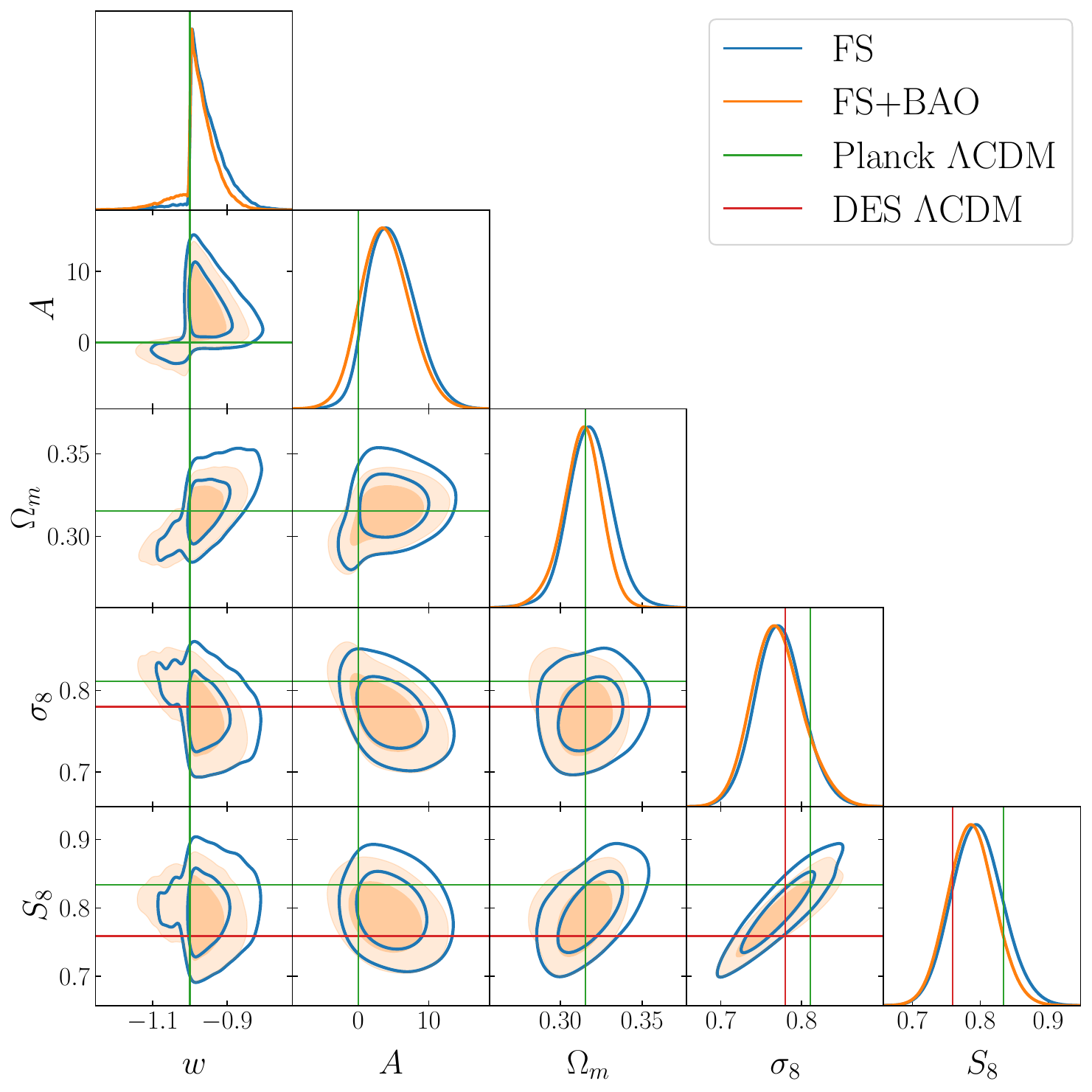}
\caption{\label{fig:baseline_BAO_OmS8} Marginalized posteriors for the dark energy parameters and the derived parameters $S_8$, $\sigma_8$ and $\Omega_m$, for the same analysis of Fig.~\ref {fig:baseline_BAO}. We also show the values of $S_8$ and $\sigma_8$ measured in the fiducial $\Lambda$CDM DES analysis~\cite{DES:2021bvc, Secco:2021vhm} with red solid lines, demonstrating our good agreement with those experiments.}
\end{figure}

The posteriors for all cosmological parameters sampled are shown in Figs.~\ref{fig:baseline_BAO} and \ref{fig:baseline_BAO_OmS8}, with the means and best-fit parameters shown in Table~\ref{tab_baseline_BAO}. We show both the case including only the full shape data (FS) as well as the full data set, including also BAO data (FS+BAO).

\begin{figure}[tbp]
\centering 
\includegraphics[width=.9\textwidth]{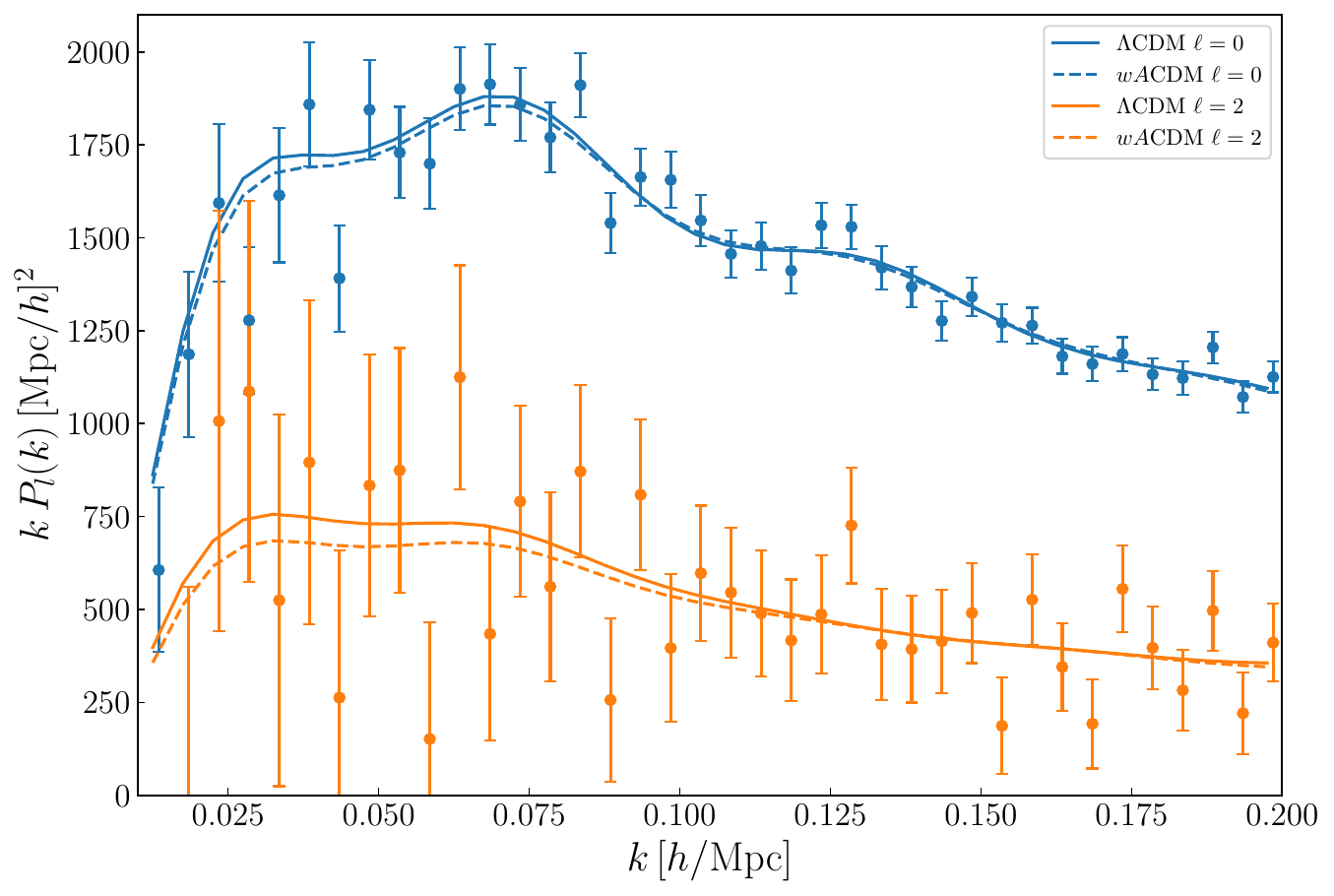}
\caption{\label{fig:comp_wA_lcdm} Comparison of the monopole (blue lines and points) and quadrupole (orange lines and points) for the best-fit point in our baseline $wA$CDM case (dashed lines), to the corresponding $\Lambda$CDM case (solid lines). The multipoles are computed with all cosmological parameters matching except for $w$ and $A$, but with marginalised nuisance parameters differing. It is this compensation from the marginalised parameters that induces this variation to affect the quadrupole substantially more than the monopole. Fixing all nuisance parameters alters the amplitude of both multipoles close to proportionally, as expected.}
\end{figure}

From this analysis we find that there is a slight preference for a positive value of $A$ at a significance of $\sim1\sigma$, with our measurement giving 
\be
A=3.9^{+3.2}_{-3.7}\ (5.4) \ {\rm b/GeV}\,,
\ee
with the best-fit value shown in parenthesis. The origin of this preference for $A>0$ is born out of the ability of the dark scattering interaction to lower the value of $\sigma_8$ relative to that predicted from $\Lambda$CDM or $w$CDM. This has the effect shown in Fig~\ref{fig:comp_wA_lcdm}: increasing $A$ while marginalizing over nuisance parameters mostly suppresses the quadrupole, while only mildly affecting the monopole. This effect of lowering $\sigma_8$ on the quadrupole had already been shown by Ref.~\cite{Philcox:2021main}, but here this is achieved with a concrete model, which additionally also lowers the growth factor $f$.

\begin{table}[h]
\centering
 \begin{tabular}{||l l l ||} 
 \hline 
 Parameter & $wA$CDM FS & $wA$CDM FS+BAO \\
 \hline\hline
 $w$ &  $-0.954^{+0.024}_{-0.046}$ ($-0.987$) & $-0.972^{+0.036}_{-0.029} $ ($-0.994$) \\ 
 \hline
 $A$ &  $4.8^{+2.8}_{-3.8} $ ($3.7$) & $3.9^{+3.2}_{-3.7} $ ($5.4$)\\ \hline
$\omega_c$ & $0.1239^{+0.0062}_{-0.0069}$ ($0.1335$) & $0.1201^{+0.0052}_{-0.0058} $ ($0.120$) \\  
 \hline
 $100 \omega_b$ & $2.267\pm 0.038$ ($2.296$) & $2.266\pm 0.038$ ($2.268$)\\ 
 \hline
 $h$ & $0.680\pm 0.012 $ ($0.6912$) & $0.677\pm 0.011$ ($0.677$)\\ 
 \hline
$\log(10^{10} A_s)$ & $3.038\pm 0.042$ ($3.060$)  & $3.039\pm 0.043$ ($3.033$)\\  
 \hline
 $n_s$ & $0.964\pm 0.012 $ ($0.958$) & $0.966\pm 0.012$ ($0.962$) \\ 
 \hline\hline
 $\Omega_m$ &  $0.318\pm 0.014 $ ($0.329$) & $0.313^{+0.013}_{-0.011}$ ($0.313$)\\ 
 \hline
 $\sigma_8$ &  $0.773^{+0.028}_{-0.031}$ ($0.839$) & $0.771^{+0.028}_{-0.034} $ ($0.762$)\\ 
 \hline
 $S_8$ &  $0.796\pm 0.037$ ($0.878$) & $0.787\pm 0.034$ ($0.778$) \\ 
 \hline
\end{tabular}
\caption{Parameter constraints for the two baseline analyses with interacting dark energy, here labelled $wA$CDM. We show results using full shape only (FS) as well as those including measurements of the BAO scale (FS+BAO). We show the 7 sampled parameters and 3 derived parameters, $S_8$, $\sigma_8$ and $\Omega_m$. The values shown are the means with the $1\sigma$ confidence intervals and the best-fit values in parenthesis.}
\label{tab_baseline_BAO}
\end{table}

Our hint of $A>0$ is therefore a translation of the preference of the BOSS data for a low $\sigma_8$, already shown above in the analyses in $\Lambda$CDM and $w$CDM. This is visible directly in Fig.~\ref{fig:baseline_BAO_OmS8}, where we additionally demonstrate that our measurement of $\sigma_8$ is in agreement with that from weak lensing surveys such as DES~\cite{DES:2021bvc, Secco:2021vhm}, and lower than that inferred from the Planck measurements with $\Lambda$CDM.

As in our previous analyses, the other cosmological parameters not fixed by their priors are fully in agreement with the values measured by Planck. A distinction with respect to the $\Lambda$CDM and $w$CDM analyses is that the measurement of $A_s$ is not being pushed as much towards lower values as in those cases, indicating that the preference of the data for a low $\sigma_8$ is being driven by the interaction, via $A$. Furthermore, given that this dark sector interaction only affects the late-time evolution and does not modify early-time physics, its effect on the CMB is negligible, as shown in Ref.~\cite{Linton:2021cgd}. It is therefore justified that we use the Planck prior on the primordial parameters, whose measurements would be unaffected by the inclusion of the interaction. This indicates that this model of interacting dark energy can alleviate the $\sigma_8$ tension, thus re-establishing the concordance between the early- and late-Universe measurements of the clustering amplitude. 

As discussed in the context of the $w$CDM analysis, including BAO data has a considerable effect in breaking the degeneracy between $w$ and $h$, and therefore allows for stronger constraints on those parameters as well as on $\omega_c$~\cite{DAmico:2020kxu,Philcox:2021main,Philcox:2020vvt,Chudaykin:2020ghx}. Despite having a small impact on the measurement of $A$, since it only probes the expansion history, it allows for a tighter constraint on $w$, and therefore alleviates the degeneracy between $w$ and $A$~\cite{Carrilho:2021hly}, which moves the bounds on $A$ towards lower values.

We show the full contours with all sampled parameters in Fig.~\ref{fig:baseline_all_pars} and summary results in Table~\ref{tab_baseline_BAO_bias} of Appendix~\ref{sec:full_cont}. It can be seen there that this result does not come at the cost of large or unusual bias parameters, with the linear bias being always $\sim2$. There are some differences, however, with respect to the bias parameters measured by Ref.~\cite{Philcox:2021main}, which detect larger values of linear bias. However, it is plausible that these would be different in a cosmology with a dark matter interaction, since this affects the dynamics of collapse, as seen for example in Ref.~\cite{Carrilho:2021rqo}.\\

\begin{figure}[tbp]
\centering 
\includegraphics[width=.9\textwidth]{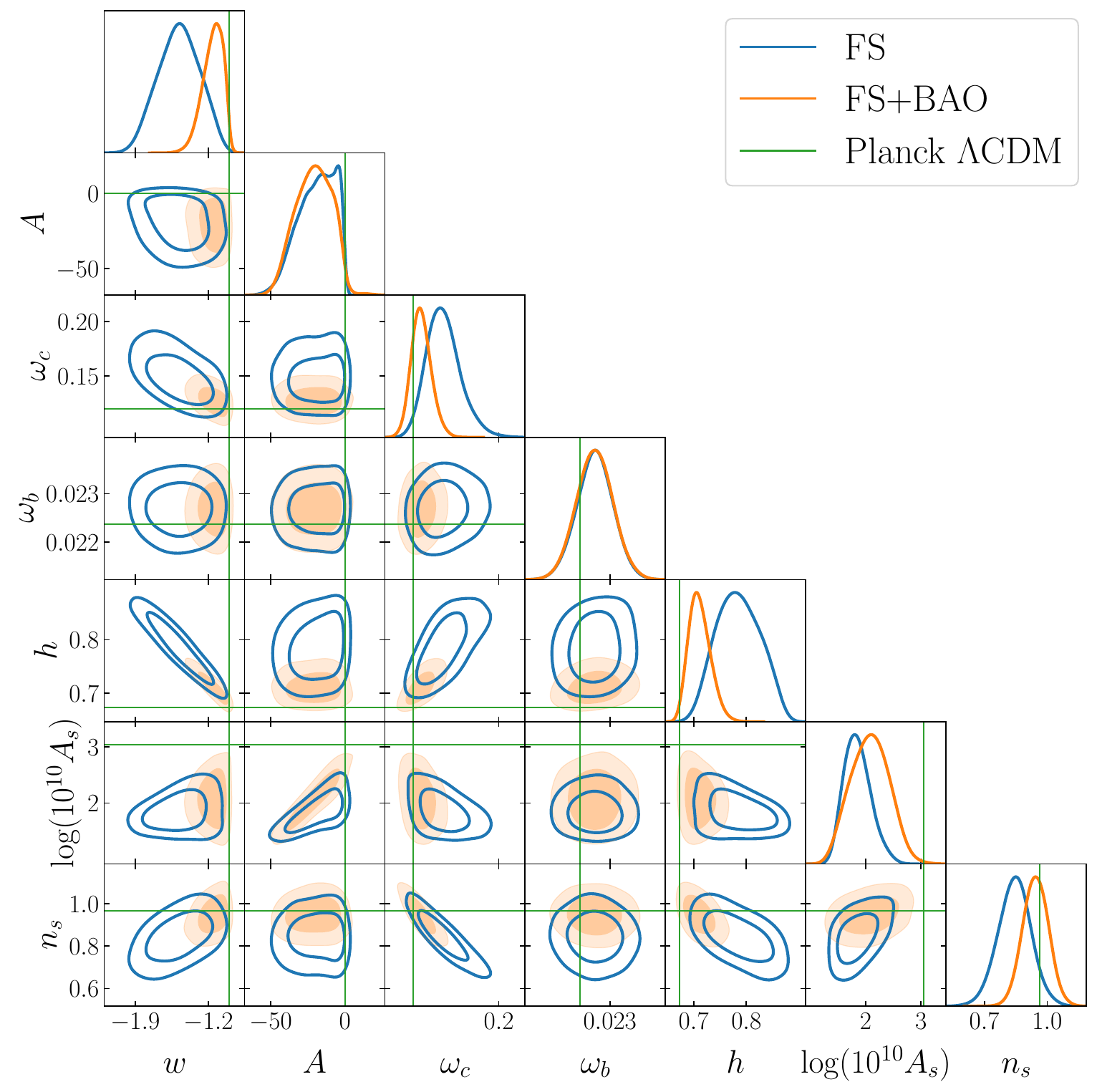}
\caption{\label{fig:cmb_free_cosmo}  Marginalized posteriors for the sampled cosmological parameters in the CMB-free analysis (with a BBN prior on $\omega_b$). We use $k_{\rm max}=0.20~h/{\rm Mpc}$ for all 3 multipoles and show both the analysis with the full shape only (FS) and with both the full shape and the BAO measurement (FS+BAO). The shape of the $w-A$ contour is due to the physical prior imposed on those parameters, i.e. ${\rm sign}(A)={\rm sign}(1+w)$.}
\end{figure}

\begin{figure}[tbp]
\centering 
\includegraphics[width=.9\textwidth]{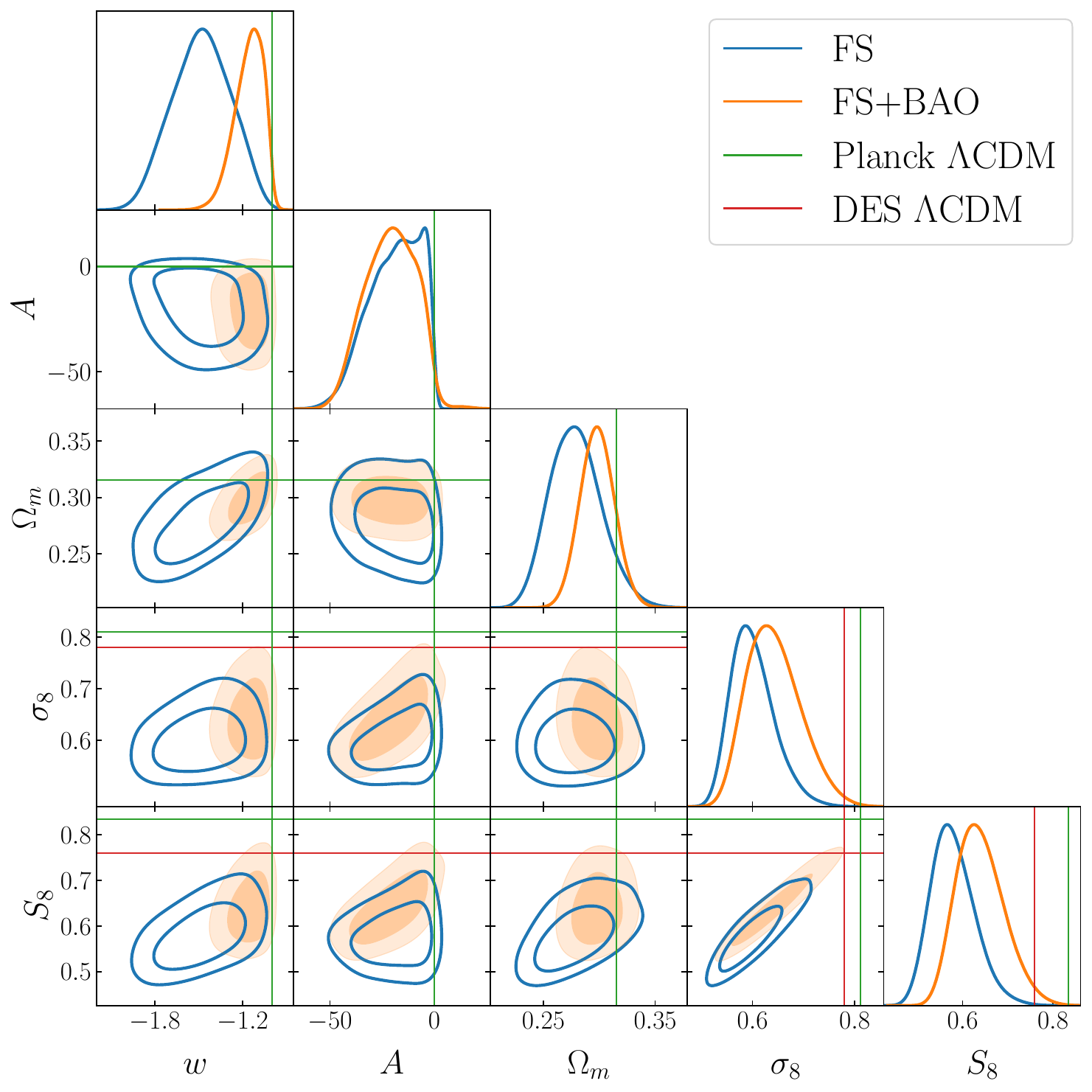}
\caption{\label{fig:cmb_free_cosmo_der} Marginalized posteriors for the dark energy parameters and the derived parameters $S_8$, $\sigma_8$ and $\Omega_m$, for the same analysis of Fig.~\ref {fig:cmb_free_cosmo}. We also show the values of $S_8$ and $\sigma_8$ measured in the fiducial $\Lambda$CDM DES analysis~\cite{DES:2021bvc, Secco:2021vhm}, now showing complete disagreement due to the degeneracy detailed in the text.}
\end{figure}

We now explore the more general CMB-free analysis. We show the posteriors of the sampled cosmological parameters in Fig.~\ref{fig:cmb_free_cosmo} and with derived parameters in Fig.~\ref{fig:cmb_free_cosmo_der}. The summary of the measured parameters can be found in Table~\ref{tab_cmb_free_BAO}. Full contours and more details can be found in Fig.~\ref{fig:cmb_free_all_pars} and Table~\ref{tab_cmb_free_BAO_bias} of Appendix~\ref{sec:full_cont}.

We can see explicitly here that without prior early Universe information, the effect of the BAO data is even stronger. Again in this case, one of the important effects of the BAO is to break the degeneracy between $w$ and $h$. Indeed, as seen in the $w$CDM analyses above and also found in Refs.~\cite{DAmico:2020kxu,Chudaykin:2020ghx}, the BAO data from various different redshifts is essential for achieving a better constraint on $w$.

\begin{table}[h]
\centering
 \begin{tabular}{||l l l ||} 
 \hline 
 Parameter & $wA$CDM FS & $wA$CDM FS+BAO \\
 \hline\hline
 $w$ &  $-1.48\pm 0.20$ ($-1.28$) & $-1.16^{+0.12}_{-0.069} $ ($-1.097$) \\ 
 \hline
 $A$ &  $-18.9^{+17}_{-7.7}  $ ($-0.7631$) & $-21^{+14}_{-11}  $ ($-0.1077$)\\ 
 \hline
$\omega_c$ & $0.148^{+0.013}_{-0.018}$ ($0.138$) & $0.1275^{+0.0082}_{-0.010} $ ($0.1249$) \\  
 \hline
 $100 \omega_b$ & $2.269\pm 0.038$ ($2.266$) & $2.268\pm 0.038        $ ($2.253$)\\ 
 \hline
 $h$ & $0.785^{+0.041}_{-0.046}$ ($0.746$) & $0.711^{+0.017}_{-0.024}$ ($ 0.6992$)\\ 
 \hline
$\log(10^{10} A_s)$ & $1.85^{+0.20}_{-0.28} $ ($2.37$)  & $2.07\pm 0.35$ ($2.725$)\\  
 \hline
 $n_s$ & $0.844\pm 0.080$ ($0.885$) & $0.945\pm 0.062$ ($0.9553$) \\ 
 \hline\hline
 $\Omega_m$ &  $0.279^{+0.020}_{-0.026} $ ($0.290$) & $0.298\pm 0.015$ ($0.303$)\\ 
 \hline
 $\sigma_8$ &  $0.601^{+0.032}_{-0.049}$ ($0.665$) & $0.642^{+0.044}_{-0.061}$ ($0.730$)\\ 
 \hline
 $S_8$ &  $0.579^{+0.040}_{-0.054}$ ($0.654$) & $0.640^{+0.044}_{-0.060}$ ($0.733$) \\
 \hline
\end{tabular}
\caption{Parameter constraints for the two CMB-free analyses with interacting dark energy, here labelled $wA$CDM. We show results using full shape only (FS) as well those including measurements of the BAO scale (FS+BAO). We show the 7 sampled parameters as well as 3 derived parameters, $S_8$, $\sigma_8$ and $\Omega_m$. The values shown are the means with the $1\sigma$ confidence intervals and the best-fit value in parenthesis.}
\label{tab_cmb_free_BAO}
\end{table}

Let us now comment on the results themselves. We can immediately see that the data now appear to prefer a negative value of $A$, as well as a very low value of the scalar amplitude, independently of the inclusion of BAO data. This is explained by a very strong degeneracy between $A$ and $A_s$, which is also clearly seen in the posterior plots. Physically, as $A$ grows more negative, the amplitude of the linear power spectrum increases and $A_s$ decreases to compensate, being even lower due to the preference of the data for a low $\sigma_8$. This is compounded by the linear bias also being largely degenerate with the amplitude. This is seen very clearly in the full contours in Fig.~\ref{fig:cmb_free_all_pars} of Appendix~\ref{sec:full_cont}, where we can also see that the linear bias is very large and is already dominated by its prior in some cases. Should that prior be relaxed, this degeneracy would likely drive the bias to even larger values, further reducing the scalar amplitude and increasing $A$. We can also see in Fig.~\ref{fig:cmb_free_cosmo_der} that the preferred value of $\sigma_8$ and $S_8$ is equivalently low, given this degeneracy with the bias. Similar degeneracies are expected to occur in other beyond-$\Lambda$CDM cosmologies and have been seen in the case of non-flat $\Lambda$CDM~\cite{Glanville:2022xes}, as well as in a very recent analysis in the context of nDGP gravity~\cite{Piga:2022mge}, both of which are likely to arise for the same reasons.

In general, the BOSS full shape data is able to constrain two combinations of these three amplitude parameters, essentially via the measurement of the amplitudes of the monopole and quadrupole. Therefore, when all three are allowed to vary this degeneracy is inevitable, whereas in the analysis with the CMB prior on $A_s$, this problem is resolved and a fairly precise measurement of $A$ and $b_1$ is possible. In principle, nonlinear effects can help resolving this degeneracy, since they are sensitive to other combinations of $A_s$, $A$ and $b_1$. This can be achieved by the enhanced precision of future stage-IV surveys and/or by improved modelling.

In spite of the degeneracy found in these analyses, there still appears to be a preference for a negative value of $A$, which remains to be explained. We believe this preference is driven by the slight preference of LSS data for $w<-1$. This is seen in the $w$CDM analyses in the literature (e.g.~\cite{DAmico:2020kxu,Chudaykin:2020ghx}) as well as our own (CMB-free) analysis above, where the phantom model is preferred. Planck also appears to find a preference for phantom dark energy \cite{Aghanim:2018eyx}. Due to the physical requirement of matching signs between $A$ and $1+w$, this naturally implies that negative $A$ is preferred, since that is determined by $w<-1$ and given that $A$ is otherwise unconstrainable due to the strong degeneracies mentioned above.

Finally, we note once more that these results are somewhat dependent on the priors imposed on the nuisance parameters. As seen in more detail in Figs.~\ref{fig:cmb_free_prior_wAcdm} and \ref{fig:cmb_prior_prior_wAcdm} and Table~\ref{tab_prior_wACDM_bias} of Appendix~\ref{sec:prior}, changes to these priors can alter cosmological parameter constraints to some degree, including the value of $A$. In the case with a CMB prior on the primordial parameters, the analyses with larger priors reduce the significance of the above hint for $A>0$ to approximately $0.5\sigma$, both because the contours are larger, but also due to a shift of parameters towards their $\Lambda$CDM values. This is complemented by a fairly unrealistic shift of $\omega_c$ to small values, which would be in $2$ to $3\sigma$ tension with Planck. However, in such a situation where the priors appear to be informative, it is important to be conservative on any claimed hint or detection.

\section{Conclusions}
\label{sec:conclusions}

We have analysed the power spectrum multipoles from the BOSS DR12 data in conjunction with BAO data from a broad range of redshifts, in the context of three different cosmological models. We use data vectors generated from the new windowless estimator for the power spectrum~\cite{Philcox:2020vbm,Philcox:2021main} and follow state-of-the-art modelling and analysis techniques using a fully independent likelihood pipeline, with a theoretical prediction based on the linear emulator \texttt{bacco}~\cite{Arico:2021izc} and the perturbation theory code \texttt{FAST-PT}~\cite{McEwen:2016,Fang:2017}. For all cosmological models, we perform two analyses, one including a Planck prior on primordial parameters $A_s$ and $n_s$ and one free of CMB information.

In the context of $\Lambda$CDM we find a very low scalar amplitude when using only the full-shape data, which is fully in agreement with Ref.~\cite{Philcox:2021main}, as can be seen in Fig.~\ref{fig:lcdm_comp_PI}. Adding the BAO data and in the CMB-free analysis, we measure a larger value of the amplitude,
\be
\log(10^{10}A_s) = 2.821 \pm 0.158\ (2.832)\,,
\ee
which is still low relative to the Planck results. The measured value is, however, in broad agreement with that found in many other full-shape analyses of the BOSS data, but is in slight disagreement with analyses including the eBOSS quasar sample and/or with template-based modelling, including the official combined analysis of Ref.~\cite{Alam:2020sor}. The origin of these differences will be investigated in future work.

In our CMB-dependent analysis, all parameters are broadly consistent with Planck, with only the amplitude being slightly pushed to low values within the prior we use, resulting in a slightly lower $\sigma_8$ than expected from Planck. See Fig.~\ref{fig:lcdm} and Table~\ref{tab_baseline_lcdm} for more details.

For $w$CDM, our CMB-free analysis shows a degeneracy between $w$ and the amplitude, which brings the amplitude to lower values than in the $\Lambda$CDM case and drives the EOS to the phantom regime:
\be
w=-1.17^{+0.12}_{-0.11}\ (-1.093)\,.
\ee
The analysis with a CMB prior, on the other hand, moves the constraint on $w$ towards the $\Lambda$CDM value and all other parameters are consistent with Planck, with the amplitude being the same as in the corresponding $\Lambda$CDM case. All these results are shown in Fig.~\ref{fig:wcdm} and Table~\ref{tab_baseline_wcdm}.

In the analysis for $wA$CDM, we find very large degeneracies between the interaction strength $A$, the scalar amplitude $A_s$ and the linear bias, in the context of the CMB-free case, which is shown in Figs.~\ref{fig:cmb_free_cosmo} and \ref{fig:cmb_free_cosmo_der}. This makes it very difficult to obtain a constraint on $A$, with a negative value being preferred, driven by the preference for $w<-1$ and the mentioned degeneracy. However, in our main analysis including a CMB prior shown in Figs.~\ref{fig:baseline_BAO} and \ref{fig:baseline_BAO_OmS8}, the degeneracies are broken and we find a hint for a positive interaction strength:
\be
w=-0.972^{+0.036}_{-0.029}\ (-0.994)\,,\quad A= 3.9^{+3.2}_{-3.7}\ (5.4)\,.
\ee
This leaves the scalar amplitude $A_s$ closer to the Planck value than in the other corresponding analyses, while also lowering $\sigma_8$ to similar values as seen in weak lensing experiments. This represents the first measurement of this parameter and shows that the Dark Scattering model can be successful in re-establishing concordance between early and late Universe measurements of the scalar amplitude. It is therefore important to study this model in more detail with different and future data, both spectroscopic and photometric, to fully understand whether it can consistently explain all observations.

We studied the dependence of all analyses on the priors of the nuisance parameters in Appendix~\ref{sec:prior}. We find in all cases that these priors are informative, as broadening them modifies the posterior distribution for cosmological parameters. For the CMB-free cases, the most affected parameter is the spectral index, $n_s$, although the amplitude $A_s$ can also be substantially affected, with other parameters less so. When a CMB prior is imposed, the effect is instead to change mostly $\omega_c$ as well as $h$ to a smaller extent. The dark energy parameters $w$ and $A$ are also shifted by fractions of their base uncertainties, typically in the direction of their $\Lambda$CDM values. This means, for example, that our main result on the strength of the interaction is less significant when the priors are allowed to vary in a larger range. \\

Finally, let us comment on the overall conclusions of this work. While we find interesting new constraints on dark energy, in all cases tested and for all models under study, varying the nuisance parameter priors leads to significant shifts in the cosmological parameter constraints, in addition to a broadening of the posteriors. This implies that the priors on the nuisance parameters contain non-negligible information, which determines the possible constraints on cosmology, at least at the level of precision of the BOSS data. The main question is whether this information is correct and how it can be obtained. As mentioned in the main text, these prior ranges were chosen based on several theoretical arguments relating to the consistency of the bias model, and based on the values typically seen in simulation studies. However, galaxies in the real Universe could be different from simulations, particularly given the difficulty of modelling galaxy formation robustly, as well as simply due to selection effects on the galaxy samples used for measuring spectra. Moreover, the theoretical arguments are typically order-of-magnitude statements, which are difficult to implement exactly. Given these results and the uncertainty in predicting values of nuisance parameters, we argue that it is crucial to study this in more depth in the future. A better understanding of how nuisance parameter priors affect constraints is crucial for confirming or ruling out future apparent tensions or the detection of exotic new physics.\\\\

After this paper was first released, another work~\cite{Simon:2022lde} appeared, which performed a thorough study of the impact of different analysis choices available in the literature. They conclude that the most impactful choice is that of the priors and obtain similar results to ours when the priors are broadened. These results confirm the importance of this issue and motivate a detailed study of the information content of priors of nuisance parameters, not only for BOSS analyses, but also in the context of stage IV surveys.

\acknowledgments

We acknowledge use of the Cuillin computing cluster of the Royal Observatory, University of Edinburgh.
AP is a UK Research and Innovation Future Leaders Fellow [grant MR/S016066/1]. PC and CM's research is supported by a UK Research and Innovation Future Leaders Fellowship [grant MR/S016066/1]. 
For the purpose of open access, the author has applied a Creative Commons Attribution (CC BY) licence to any Author Accepted Manuscript version arising from this submission.

\bibliographystyle{JHEPmodplain}
\bibliography{IDE_biblio}

\appendix

\section{Dependence on priors of nuisance parameters}
\label{sec:prior}

In this appendix we study the prior dependence of our results. We focus on the nuisance parameters, namely the bias parameters $b_2$, $b_{\mathcal{G}_2}$ and $b_{\Gamma_3}$, the counter-term parameters $c_0$, $c_2$, $c_4$ and $c_{\nabla^4\delta}$, and the shot-noise parameters 
$N$, $e_0$ and $e_2$. We repeat the main FS+BAO analyses of the main text, but modify the standard deviation of the Gaussian prior of these nuisance parameters\footnote{We do not modify the prior on $b_1$, since the range chosen of $[0,4]$ is already very broad.} by factors of 3 and 10, denoting those cases by $3\times\sigma$ and $10\times\sigma$, respectively. The finding of different constraints on the cosmological parameters in this exercise when compared to the baseline cases implies that these priors are informative and therefore need to be chosen very carefully.

\subsection{$\Lambda$CDM}

\begin{table}[h]
\fontsize{8.5}{10}\selectfont
\centering
 \begin{tabular}{||l| l l | l l ||} 
 \hline 
 Parameter & $3\times\sigma$ CMB-free & $10\times\sigma$ CMB-free & $3\times\sigma$ CMB prior & $10\times\sigma$ CMB prior\\
 \hline\hline
 $\omega_c$ & $0.124^{+0.009}_{-0.011}$ ($0.123$) & $0.111^{+0.007}_{-0.010} $ ($0.111$) &  $0.108\pm 0.005$ ($0.107$) & $0.100\pm 0.006$ ($0.096$) \\  
 \hline
 $100 \omega_b$ & $2.267\pm 0.038 $ ($2.275$) & $2.267\pm 0.038 $ ($2.279$)&  $2.263\pm 0.038$ ($2.245$)& $2.264\pm 0.038$ ($2.246$)\\ 
 \hline
 $h$ & $0.685\pm 0.010  $ ($0.685$) & $0.673^{+0.009}_{-0.011}$ ($0.676$) &  $0.671\pm 0.008$ ($0.672$) & $0.665\pm 0.009  $ ($0.661$)\\ 
 \hline
$\log(10^{10} A_s)$ & $2.69\pm 0.18$  ($2.76$) & $2.22^{+0.15}_{-0.24} $ ($2.14$)&  $3.029\pm 0.040 $ ($3.034$)& $3.028\pm 0.041  $ ($3.048$)\\  
 \hline
 $n_s$ & $0.810\pm 0.083 $  ($0.814$)& $0.029^{+0.066}_{-0.19}$ ($-0.102$) &  $0.963\pm 0.012$ ($0.966$)& $0.961\pm 0.012  $ ($0.955$)\\ 
 \hline\hline
 $b_1^{Nz1}$ &  $2.10\pm 0.24  $ ($2.01$) & $1.14^{+0.23}_{-0.28}$ ($1.02$)&  $1.91\pm 0.09$ ($1.91$)& $1.89\pm 0.14 $ ($1.847$)\\ 
 \hline
 $b_2^{Nz1}$ &  $-1.4^{+1.9}_{-3.0}$ ($-2.0$)& $-4.6\pm 6.7 $ ($-7.9$)&  $-1.1^{+1.5}_{-2.8}$ ($-1.8$)& $-0.9\pm 4.6 $ ($-2.4$)\\ 
 \hline
 $b_{\mathcal{G}_2}^{Nz1}$ &  $-1.3\pm 1.4$ ($-1.2$)& $-9.5\pm 4.1 $ ($-12.0$)&  $-0.5\pm 1.2 $ ($-0.4$)& $-0.4^{+3.3}_{-2.8} $ ($-1.5$)\\ 
 \hline
 $b_1^{Sz1}$ &  $2.17\pm 0.23  $ ($2.14$)& $1.21^{+0.24}_{-0.29}   $ ($1.13$)&  $1.98^{+0.11}_{-0.10}  $ ($2.01$)& $1.90^{+0.20}_{-0.14}$ ($1.83$)\\ 
 \hline
 $b_2^{Sz1}$ &  $-0.4^{+2.4}_{-3.3} $ ($1.0$) & $-1.2\pm 8.2$ ($1.6$)&  $0.0^{+2.1}_{-3.2} $ ($-1.2$) & $0.2^{+9.1}_{-8.2}  $ ($-5.4$)\\ 
 \hline
 $b_{\mathcal{G}_2}^{Sz1}$ &  $-0.2\pm 1.3 $ ($0.5$) & $-7.4\pm 4.7 $ ($-7.6$) &  $0.2\pm 1.2$ ($0.1$)& $0.9^{+3.2}_{-2.8} $ ($-0.1$)\\ 
 \hline
 $b_1^{Nz3}$ &  $2.00\pm 0.28 $ ($1.86$)& $0.99^{+0.24}_{-0.30}  $ ($0.83$)&  $1.82^{+0.11}_{-0.10}$ ($1.83$)& $1.77^{+0.16}_{-0.13}  $ ($1.64$)\\ 
 \hline
 $b_2^{Nz3}$ &  $-5.9^{+1.8}_{-2.6}$ ($-6.4$)& $-4.5\pm 7.4  $ ($-3.3$)&  $-5.1^{+1.3}_{-1.9}$ ($-5.4$)& $-5.2^{+1.4}_{-4.2}$ ($-7.2$)\\ 
 \hline
 $b_{\mathcal{G}_2}^{Nz3}$ &  $-2.1^{+1.7}_{-1.6}$ ($-2.7$)& $-13.5^{+4.2}_{-4.9}$ ($-14.0$)&  $-0.9\pm 1.3  $ ($-1.5$)& $-1.3^{+3.3}_{-2.8} $ ($-1.7$)\\ 
 \hline
 $b_1^{Sz3}$ &  $2.40\pm 0.25   $ ($2.27$) & $1.40^{+0.26}_{-0.31}$ ($1.27$)&  $2.17^{+0.11}_{-0.10}$ ($2.16$)& $2.15^{+0.18}_{-0.15}  $ ($2.07$)\\ 
 \hline
 $b_2^{Sz3}$ &  $-0.5^{+2.7}_{-3.4} $ ($-3.0$)& $-1.3\pm 9.1 $ ($-5.5$)&  $-0.8^{+2.2}_{-3.3} $ ($-1.9$)& $-3.5^{+2.1}_{-7.2}   $ ($-7.3$)\\ 
 \hline
 $b_{\mathcal{G}_2}^{Sz3}$ &  $-1.0\pm 1.5 $ ($-1.2$)& $-8.1^{+5.0}_{-5.9}$ ($-13.1$)&  $0.0\pm 1.3 $ ($0.2$)& $1.6^{+3.3}_{-2.9} $ ($0.3$)\\ 
 \hline
\end{tabular}
\caption{Parameter constraints for all analyses with broader priors of nuisance parameters for $\Lambda$CDM. We show results using full shape + BAO data both for CMB-free analyses and those with a Planck prior. We show the 5 sampled cosmological parameters as well as 3 bias parameters per redshift bin and sky cut. The values shown are the means with the $1\sigma$ confidence intervals with the best-fit values in parenthesis.}
\label{tab_prior_LCDM_bias}
\end{table}

\begin{figure}[h]
\centering 
\includegraphics[width=\textwidth]{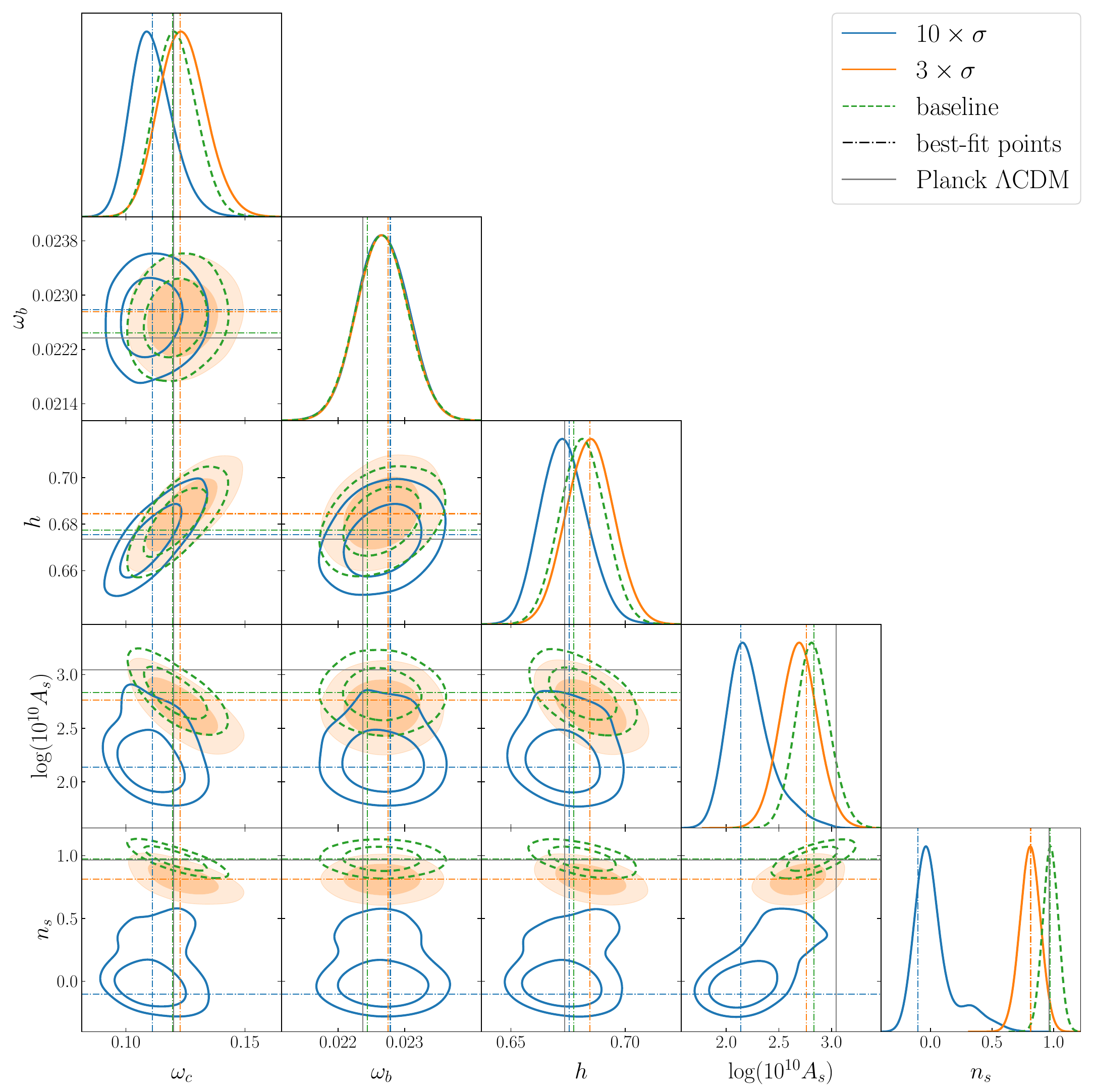}
\caption{\label{fig:cmb_free_prior_lcdm} Posteriors for the sampled cosmological parameters in the CMB-free analysis in $\Lambda$CDM for three different options for the priors on nuisance parameters: baseline case of  Eq.~\eqref{eq:prior_base} and the cases with standard deviations of nuisance parameters, except for $b_1$, increased by factors of $3$ and $10$. Dot-dashed lines show the best-fit points computed with the analytically marginalised posterior, showing that those are also shifted.} 
\end{figure}

\begin{figure}[h]
\centering 
\includegraphics[width=\textwidth]{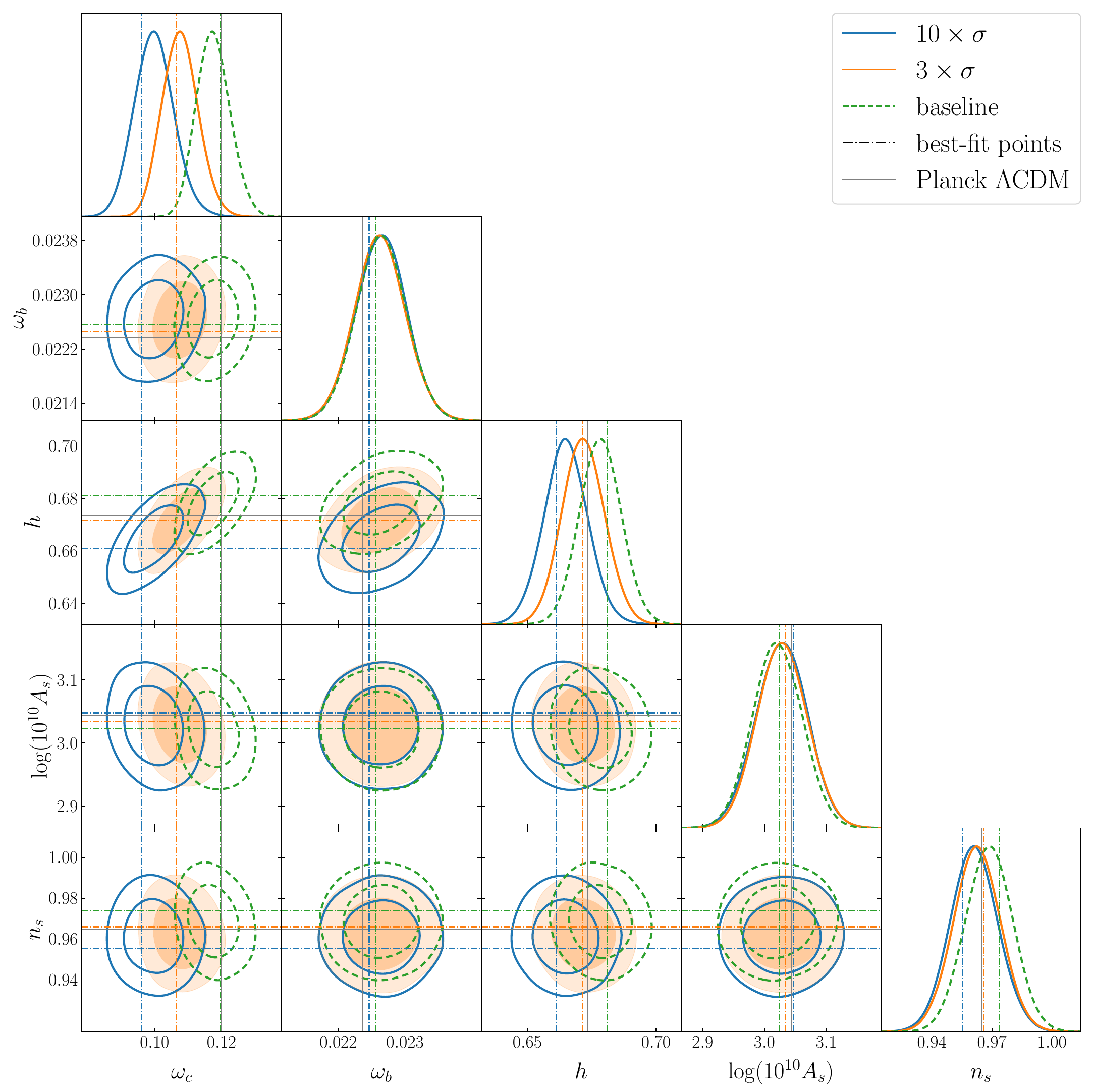}
\caption{\label{fig:cmb_prior_prior_lcdm} Posteriors for the sampled cosmological parameters in the analysis with a CMB prior in $\Lambda$CDM for three different options for the priors on nuisance parameters: baseline case of  Eq.~\eqref{eq:prior_base} and the cases with standard deviations of nuisance parameters, except for $b_1$, increased by factors of $3$ and $10$. Dot-dashed lines show the best-fit points computed with the analytically marginalised posterior, showing that those are also shifted.} 
\end{figure}

We show the results of analysing the FS+BAO data using broader priors in Fig.~\ref{fig:cmb_free_prior_lcdm} for the CMB-free case and in Fig.~\ref{fig:cmb_prior_prior_lcdm} for the case with a CMB prior on $n_s$ and $A_s$. There are some effects of altering the prior range, implying that the priors do include information. Allowing the nuisance parameters to vary an order of magnitude more than our baseline case does have a large effect, particularly on $n_s$ for the CMB-free case or on $\omega_c$ for the case with a CMB prior. There appears to be sufficient freedom in the BOSS data that the bias parameters can mimic the effects of cosmological parameters that change the shape of power spectra. In addition to this, there is also some change in the amplitude, with lower values found in the analyses with larger prior ranges.

All of these changes are, however, dependent on allowing large values for the nuisance parameters, particularly the nonlinear bias, $b_2$ and $b_{\mathcal{G}_2}$, as can be seen in Table~\ref{tab_prior_LCDM_bias}. In the CMB-free analysis and for the $10\times\sigma$ case, the preferred values of $b_{\mathcal{G}_2}$ are all close to $-10$ and the values of $b_2$ approach $-5$. Additionally, the counter-term parameters are larger than expected, with $c_0$ and $c_2$ being $O(-100)~\text{Mpc}^2h^{-2}$, and reaching $-300~\text{Mpc}^2h^{-2}$ for the low redshift bin. The $3\times\sigma$ case is less severe, but such shifts in parameters also rely on large values of the bias parameters, with $b^{Nz3}_2\sim-6$ and counter-term parameter $c_0^{Nz3}\sim100~\text{Mpc}^2h^{-2}$. 

For the analysis with a CMB prior, the effects of enlarging the prior are smaller and are transferred away from the primordial parameters into $\omega_c$. Once more, the nuisance parameters are pushed to large absolute values, with $b_2\sim-5$ and $c_0\sim100~\text{Mpc}^2h^{-2}$ at the $z=0.61$ bin, for the $10\times\sigma$ case. In the $3\times\sigma$ case only the biases have large values, with $b^{Nz3}_2\sim-5$ and this is what drives the cosmological parameters to shift considerably relative to the base case.

\subsection{$w$CDM}

\begin{table}[h]
\fontsize{8.5}{10}\selectfont
\centering
 \begin{tabular}{||l| l l | l l ||} 
 \hline 
 Parameter & $3\times\sigma$ CMB-free & $10\times\sigma$ CMB-free & $3\times\sigma$ CMB prior & $10\times\sigma$ CMB prior \\
 \hline\hline
 $w$ & $-1.20^{+0.13}_{-0.12}$  ($ -1.13$)& $-1.11^{+0.19}_{-0.12} $ ($ -0.98$)&  $-1.01^{+0.09}_{-0.08}$ ($ -0.98$) & $-1.03^{+0.10}_{-0.08}$ ($ -1.02$) \\  
 \hline
 $\omega_c$ & $0.134\pm 0.012$ ($ 0.129 $) & $0.119^{+0.011}_{-0.018} $ ($0.103 $)&  $0.108^{+0.005}_{-0.006}$ ($ 0.105$)& $0.100^{+0.006}_{-0.007}$  ($ 0.099$)\\  
 \hline
 $100 \omega_b$ & $2.266\pm 0.038 $ ($2.245 $)& $2.266\pm 0.039 $ ($2.267 $)&  $2.263\pm 0.038 $ ($ 2.242$) & $2.261\pm 0.037$ ($2.249 $) \\ 
 \hline
 $h$ & $0.723\pm 0.026   $ ($0.705 $) & $0.697^{+0.027}_{-0.042}$ ($ 0.662$) &  $0.674^{+0.015}_{-0.017}$ ($ 0.666$) & $0.670^{+0.017}_{-0.019}  $ ($ 0.668$) \\ 
 \hline
$\log(10^{10} A_s)$ & $2.46\pm 0.22$  ($ 2.63$) & $2.18^{+0.19}_{-0.23} $($ 2.23$) &  $3.030\pm 0.041 $  ($ 3.042$) & $3.025\pm 0.042 $ ($3.020 $) \\  
 \hline
 $n_s$ & $0.797\pm 0.078 $   ($0.775 $) & $0.15^{+0.42}_{-0.33}$   ($ -0.06$) &  $0.963\pm 0.012$  ($0.965 $) & $0.962\pm 0.012  $ ($0.960 $) \\ 
 \hline\hline
 $b_1^{Nz1}$ &  $2.31\pm 0.27 $   ($2.12 $) & $1.42^{+0.25}_{-0.68} $  ($ 1.01$) &  $1.90\pm 0.10$  ($ 1.88$) & $1.88\pm 0.14$ ($1.82 $) \\ 
 \hline
 $b_2^{Nz1}$ &  $-1.4^{+2.0}_{-2.8}$  ($-1.9 $) & $-4.1\pm 6.7 $  ($ -7.4$) &  $-1.3^{+1.4}_{-2.7}$  ($ -2.4$) & $0.9\pm 6.0$ ($-2.3 $) \\ 
 \hline
 $b_{\mathcal{G}_2}^{Nz1}$ &  $-1.1\pm 1.4$  ($ -1.1$) & $-8.2^{+4.4}_{-5.0} $  ($ -12.0$) &  $-0.5\pm 1.2 $  ($ -0.8$) & $-0.3^{+3.3}_{-2.8}$ ($ -0.1$) \\ 
 \hline
 $b_1^{Sz1}$ &  $2.36\pm 0.26  $  ($ 2.16$) & $1.47^{+0.28}_{-0.65}   $  ($ 1.06$) &  $1.98\pm 0.11   $  ($2.04 $) & $1.90^{+0.20}_{-0.16}$ ($ 1.89$) \\ 
 \hline
 $b_2^{Sz1}$ &  $-0.3^{+2.4}_{-3.0}  $   ($-1.0 $) & $-2.2^{+8.1}_{-9.1}$ ($ 1.0$) &  $-0.1^{+2.1}_{-3.1} $  ($-0.4 $) & $0.2^{+9.1}_{-8.2}  $ ($ -5.2$) \\ 
 \hline
 $b_{\mathcal{G}_2}^{Sz1}$ &  $0.0\pm 1.4  $  ($ -0.9$) & $-6.1\pm 5.2  $  ($ -7.7$) &  $0.2\pm 1.2$  ($ 0.2$) & $1.1^{+3.2}_{-2.9}$ ($0.4 $) \\ 
 \hline
 $b_1^{Nz3}$ &  $2.29\pm 0.34 $  ($1.96 $) & $1.28^{+0.26}_{-0.66}  $  ($0.77 $) &  $1.82\pm 0.11$  ($ 1.78$) & $1.77^{+0.16}_{-0.13}$ ($ 1.78$) \\ 
 \hline
 $b_2^{Nz3}$ &  $-5.6^{+2.0}_{-2.9}$  ($ -6.8$) & $-5.8^{+7.4}_{-9.2}   $ ($-0.5 $) &  $-5.2^{+1.3}_{-1.8}$  ($ -6.1$) & $-4.9^{+1.2}_{-4.7}$ ($ -6.3$) \\ 
 \hline
 $b_{\mathcal{G}_2}^{Nz3}$ &  $-1.8\pm 1.6 $  ($ -2.4$) & $-12.4^{+4.7}_{-5.2}$  ($ -12.3$) &  $-0.9\pm 1.3  $  ($ -0.5$) & $-1.0^{+3.2}_{-2.8}  $ ($ 0.0$) \\ 
 \hline
 $b_1^{Sz3}$ &  $2.64\pm 0.30    $ ($ 2.38$) & $1.71^{+0.33}_{-0.73}$  ($ 1.24$) &  $2.17\pm 0.11 $  ($ 2.20$) & $2.14^{+0.18}_{-0.16}  $ ($ 2.08$) \\ 
 \hline
 $b_2^{Sz3}$ &  $-0.3^{+2.8}_{-3.2} $ ($ -0.3$) & $-1.7\pm 9.2 $ ($-2.1 $) &  $-0.7^{+2.2}_{-3.4} $  ($-1.1 $) & $-3.7^{+2.1}_{-7.0} $ ($ -7.1$) \\ 
 \hline
 $b_{\mathcal{G}_2}^{Sz3}$ &  $-0.8\pm 1.5 $  ($-1.2 $) & $-7.1\pm 5.7$  ($-10.1 $) &  $0.1\pm 1.3  $  ($0.8 $) & $1.5^{+3.2}_{-2.9}$ ($ 1.3$) \\ 
 \hline
\end{tabular}
\caption{Parameter constraints for all analyses with broader priors of nuisance parameters for $w$CDM. We show results using full shape + BAO data both for CMB-free analyses and those with a Planck prior. We show the 6 sampled cosmological parameters as well as 3 bias parameters per redshift bin and sky cut. The values shown are the means with the $1\sigma$ confidence intervals with the best-fit values in parenthesis.}
\label{tab_prior_wCDM_bias}
\end{table}

\begin{figure}[h]
\centering 
\includegraphics[width=\textwidth]{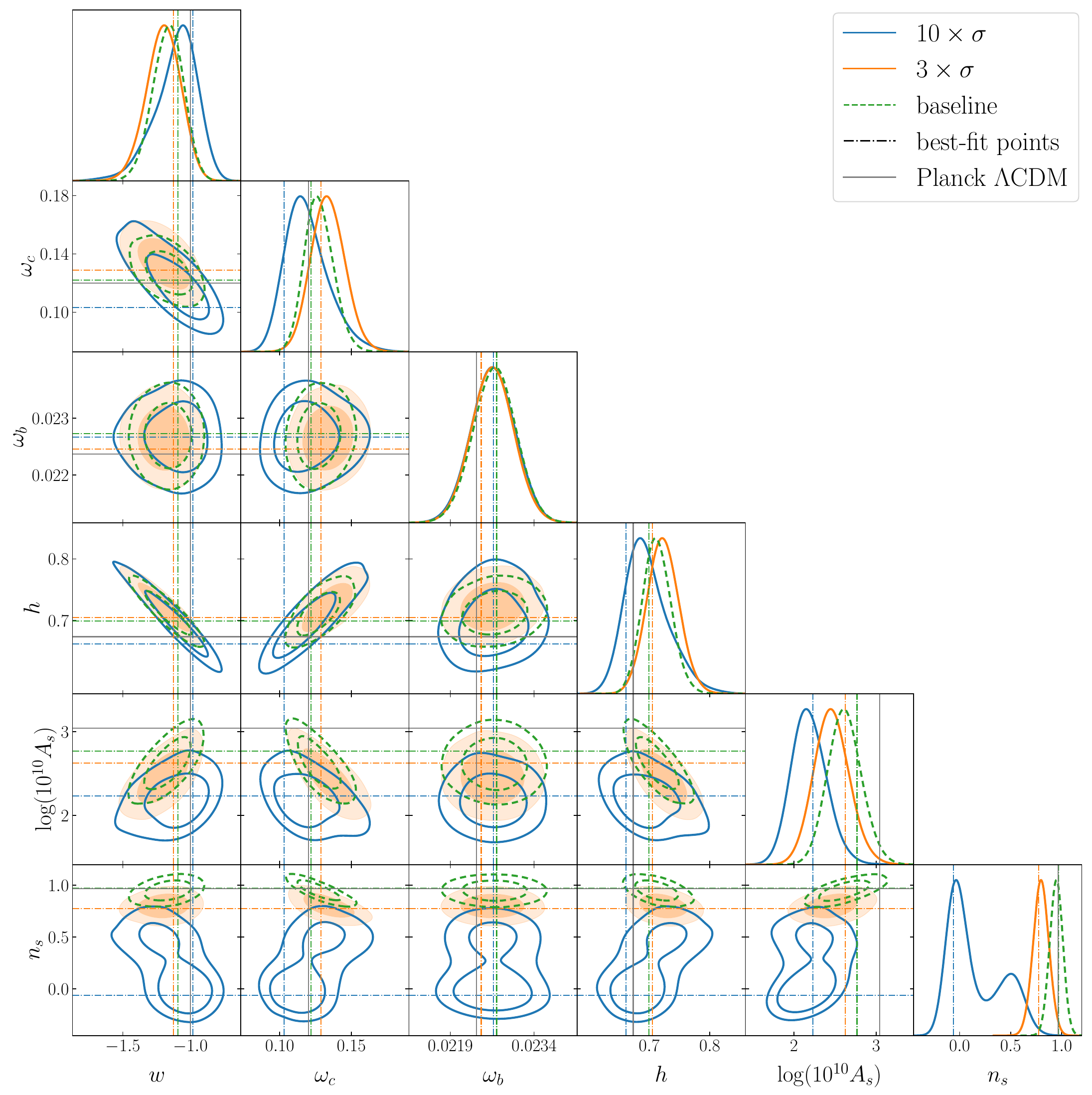}
\caption{\label{fig:cmb_free_prior_wcdm} Posteriors for the sampled cosmological parameters in the CMB-free analysis in $w$CDM for three different options for the priors on nuisance parameters: baseline case of  Eq.~\eqref{eq:prior_base} and the cases with standard deviations of nuisance parameters, except for $b_1$, increased by factors of $3$ and $10$. Dot-dashed lines show the best-fit points computed with the analytically marginalised posterior, showing that those are also shifted.} 
\end{figure}

\begin{figure}[h]
\centering 
\includegraphics[width=\textwidth]{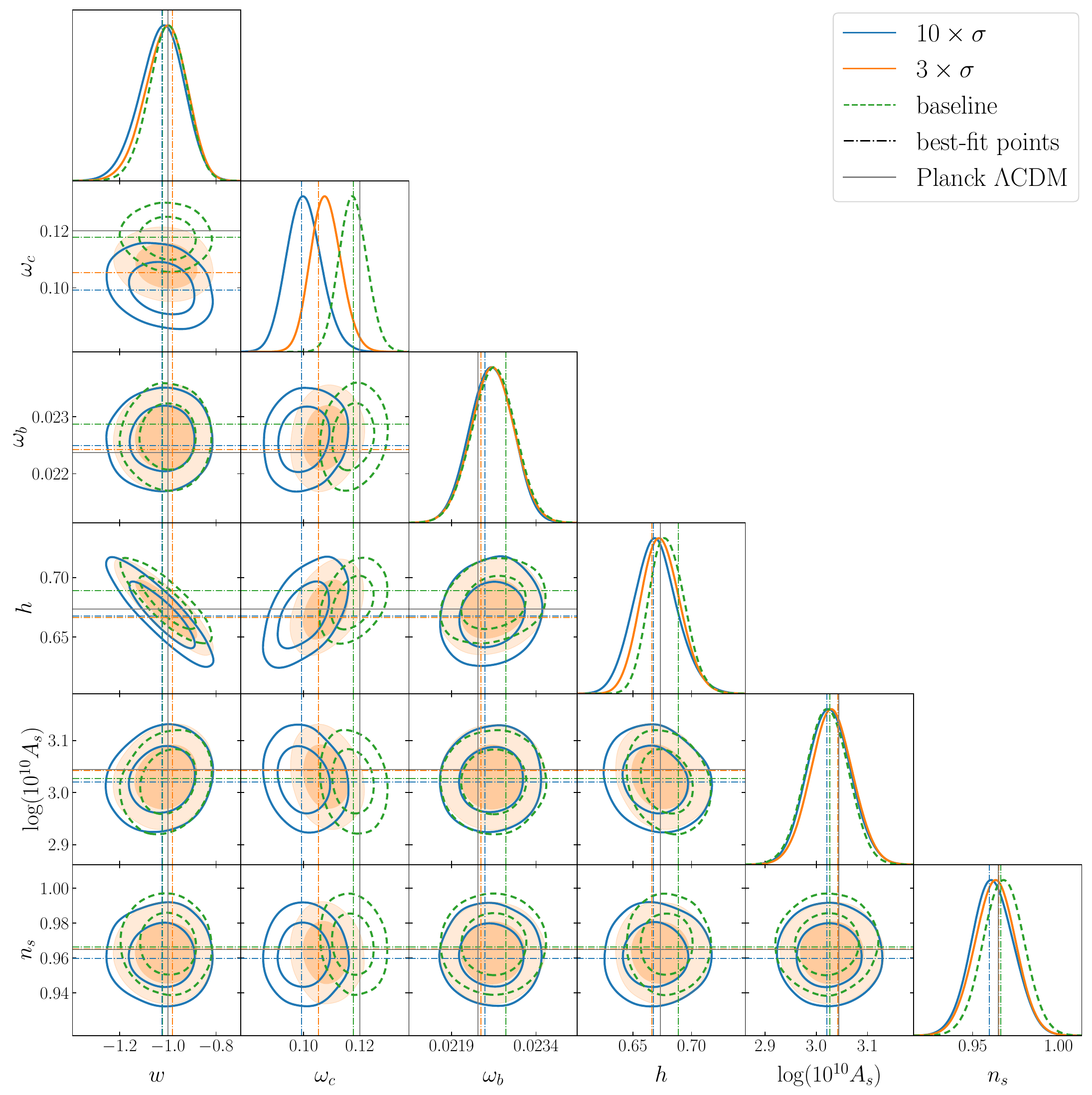}
\caption{\label{fig:cmb_prior_prior_wcdm} Posteriors for the sampled cosmological parameters in the analysis with a CMB prior in $w$CDM for three different options for the priors on nuisance parameters: baseline case of  Eq.~\eqref{eq:prior_base} and the cases with standard deviations of nuisance parameters, except for $b_1$, increased by factors of $3$ and $10$. Dot-dashed lines show the best-fit points computed with the analytically marginalised posterior, showing that those are also shifted.} 
\end{figure}

The effect of broader priors for the $w$CDM analysis is shown in Fig.~\ref{fig:cmb_free_prior_wcdm} for the CMB-free case and in Fig.~\ref{fig:cmb_prior_prior_wcdm} for the case with the prior on $A_s$ and $n_s$. Table~\ref{tab_prior_wCDM_bias} includes the constraints on all sampled parameters for all 4 different analyses, including 3 bias parameters. The conclusions of this exercise are very similar to those of the $\Lambda$CDM case, with large effects in the CMB-free and $10\times\sigma$ case, specially for $n_s$, but also for $A_s$; and smaller effects in the $3\times\sigma$ scenario. This is again driven by large values of the nonlinear bias and counter-term parameters, with the former reaching values of $O(-10)$ and the latter of $O(-300)$ in the $10\times\sigma$ case. 

In these analyses with varying $w$, that parameter is also shifted slightly when priors are modified. This small change occurs mostly in the CMB-free case when the priors are the broadest, moving $w$ towards $-1$ and increasing the size of its posterior. In the CMB-dependent analyses, only a minor broadening of the posterior is seen, with no significant shift in $w$. As in the $\Lambda$CDM case, the main effect of broadening the priors in the scenario with CMB priors is to shift the matter density measurement towards lower values.

\subsection{$wA$CDM}

\begin{figure}[h]
\centering 
\includegraphics[width=\textwidth]{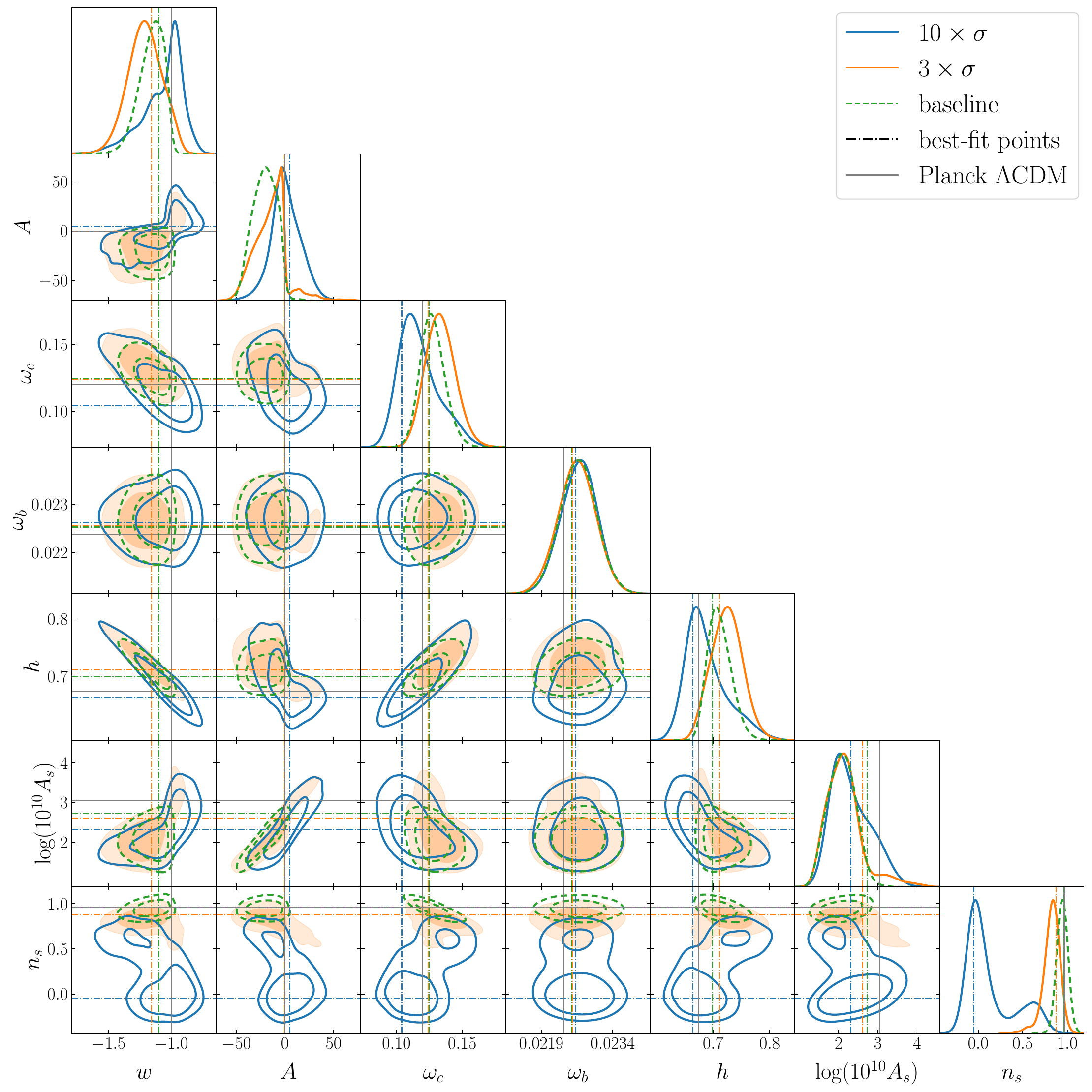}
\caption{\label{fig:cmb_free_prior_wAcdm} Posteriors for the sampled cosmological parameters in the CMB-free analysis in $wA$CDM for three different options for the priors on nuisance parameters: baseline case of  Eq.~\eqref{eq:prior_base} and the cases with standard deviations of nuisance parameters, except for $b_1$, increased by factors of $3$ and $10$. Dot-dashed lines show the best-fit points computed with the analytically marginalised posterior, showing that those are also shifted.} 
\end{figure}

\begin{figure}[h]
\centering 
\includegraphics[width=\textwidth]{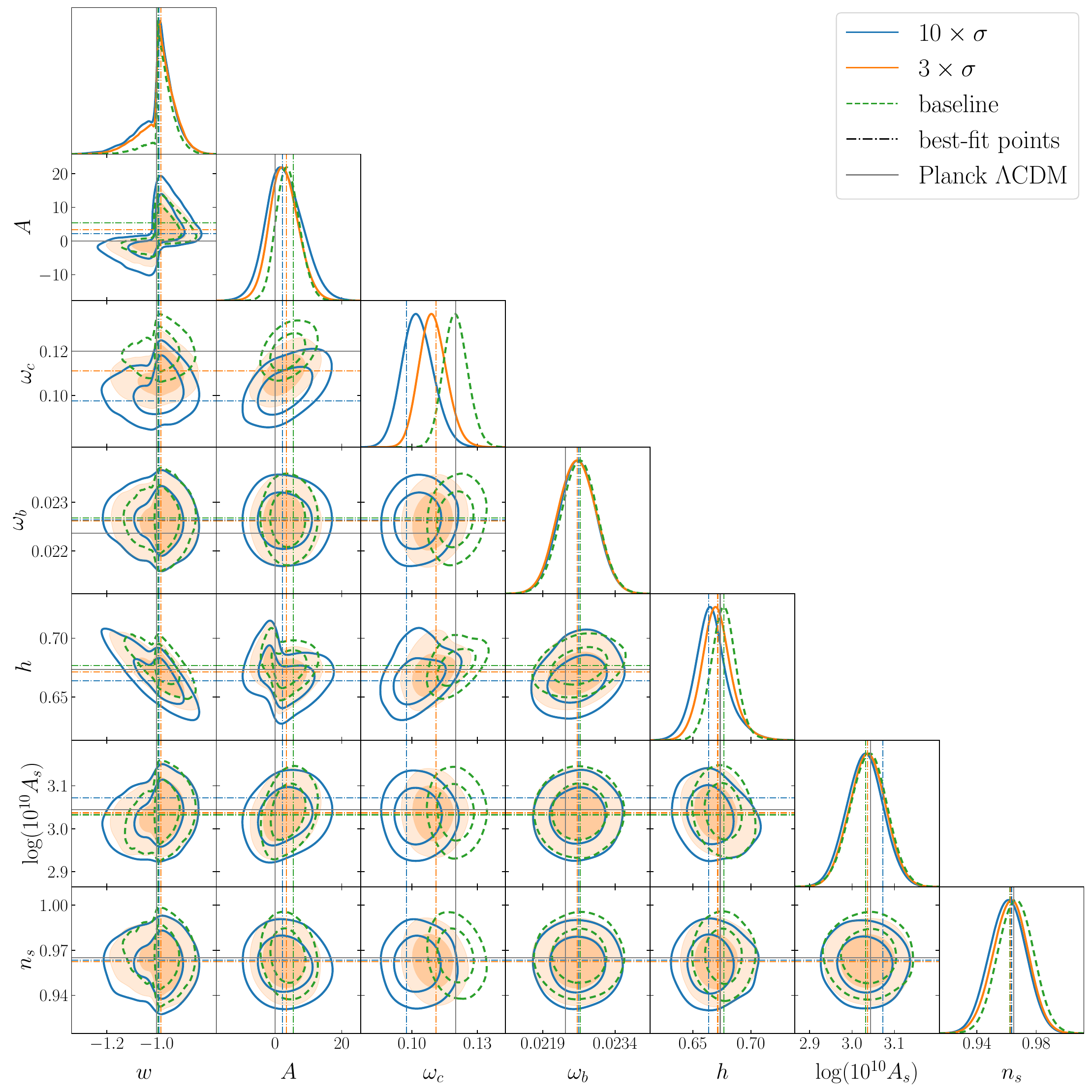}
\caption{\label{fig:cmb_prior_prior_wAcdm} Posteriors for the sampled cosmological parameters in the analysis  with a CMB prior in $wA$CDM for three different options for the priors on nuisance parameters: baseline case of  Eq.~\eqref{eq:prior_base} and the cases with standard deviations of nuisance parameters, except for $b_1$, increased by factors of $3$ and $10$. Dot-dashed lines show the best-fit points computed with the analytically marginalised posterior, showing that those are also shifted.} 
\end{figure}

The effect of larger priors in $wA$CDM is again similar to the two cases above, with a few exceptions which we will discuss. This is shown in Fig.~\ref{fig:cmb_free_prior_wAcdm} for the CMB-free case and in Fig.~\ref{fig:cmb_prior_prior_wAcdm} for the CMB-dependent case, with a summary of constraints in Table~\ref{tab_prior_wACDM_bias}.

Starting with the CMB-free case, we can see that the effects of the $10\times\sigma$ case are large, with an extreme bias on $n_s$ and smaller shifts on most other parameters. The $3\times\sigma$ case shows much milder shifts, but still substantial fractions of the parameter uncertainties. Interestingly, the preferred value of $A_s$ is not shifted, always remaining very low, even if its contours are broadened somewhat towards higher values. The consistently low value of this parameter is likely due to the degeneracy with the interaction parameter $A$, which is instead shifted towards zero as the priors are broadened. So, while $A_s$ remains fixed, a change in $A$ towards less negative values reduces the overall amplitude of the matter power spectrum, which is the same effect seen when enlarging the priors in $\Lambda$CDM and $w$CDM. 

The CMB-dependent case is also similar to the corresponding $\Lambda$CDM and $w$CDM analyses. The biggest shifts are once more in the value of $\omega_c$, but there are also some changes in $w$ and $A$. For both dark energy parameters, the effect is to increase the contours and to move them towards the $\Lambda$CDM values of $w=-1$, $A=0$. The two cases with broader priors are mostly consistent with each other, even when considering the bias parameters, as can be seen in Table~\ref{tab_prior_wACDM_bias}. Looking at the dark energy parameters only, these analyses are also consistent with the baseline case shown in the main text, but reduce the preference for interacting dark energy ($A\neq0$) to $0.5\sigma$.

\begin{table}[h]
\fontsize{8.5}{10}\selectfont
\centering
 \begin{tabular}{||l| l l | l l ||} 
 \hline 
 Parameter & $3\times\sigma$ CMB-free & $10\times\sigma$ CMB-free & $3\times\sigma$ CMB prior & $10\times\sigma$ CMB prior\\
 \hline\hline
 $w$ &  $-1.21\pm 0.14$  ($ -1.16$) & $-1.07^{+0.19}_{-0.10} $  ($ -1.00$) &  $-0.985^{+0.081}_{-0.038}$  ($-0.982 $) & $-0.994^{+0.083}_{-0.046} $  ($ -0.991$) \\ 
 \hline
 $A$ &  $-15^{+14}_{-13}$  ($-1 $) & $1\pm 16  $ ($ 5$) &  $2.7^{+3.9}_{-4.5}$  ($ 3.4$) & $3.0^{+4.9}_{-5.9}  $ ($2.2 $) \\ 
 \hline
$\omega_c$ & $0.133^{+0.011}_{-0.012}$   ($ 0.124$) & $0.117^{+0.010}_{-0.018} $  ($0.104 $) &  $0.110^{+0.006}_{-0.007}$  ($ 0.111$) & $0.103^{+0.007}_{-0.008}$   ($ 0.098 $) \\  
 \hline
 $100 \omega_b$ & $2.265\pm 0.039 $  ($ 2.256$) & $2.269\pm 0.038 $ ($2.263 $) &  $2.261\pm 0.039 $  ($ 2.261$) & $2.262\pm 0.038  $ ($2.264 $) \\ 
 \hline
 $h$ & $0.724\pm 0.028  $  ($0.711 $) & $0.688^{+0.022}_{-0.044}$  ($ 0.665$) &  $0.671^{+0.011}_{-0.013}$  ($0.672 $) & $0.666^{+0.012}_{-0.015}  $ ($ 0.664$) \\ 
 \hline
$\log(10^{10} A_s)$ & $2.15^{+0.29}_{-0.49}$   ($2.61 $) & $2.31^{+0.37}_{-0.63}$ ($2.31 $) &  $3.037\pm 0.042 $  ($3.037 $) & $3.030\pm 0.043  $ ($  3.072$) \\  
 \hline
 $n_s$ & $0.824^{+0.098}_{-0.070} $   ($0.875 $) & $0.126^{+0.059}_{-0.32}$   ($-0.049 $) &  $0.963\pm 0.012$  ($0.962 $) & $0.961\pm 0.012  $ ($ 0.963$) \\ 
 \hline\hline
 $b_1^{Nz1}$ &  $2.58^{+0.50}_{-0.32} $   ($2.19 $) & $1.32^{+0.16}_{-0.69}$  ($1.04 $) &  $1.97\pm 0.14$  ($ 1.96$) & $1.98\pm 0.21 $ ($ 1.89$) \\ 
 \hline
 $b_2^{Nz1}$ &  $-1.2^{+2.2}_{-2.9}$  ($-0.9 $) & $-4.7^{+5.9}_{-6.7}$  ($0.3 $) &  $-1.0^{+1.6}_{-2.8}$  ($ -1.4$) & $-1.1^{+3.2}_{-5.9}  $ ($ -3.1$) \\ 
 \hline
 $b_{\mathcal{G}_2}^{Nz1}$ &  $-0.9\pm 1.5$  ($-0.7 $) & $-8.1^{+4.2}_{-4.8} $  ($ -7.4$) &  $-0.5\pm 1.2 $  ($-0.7 $) & $-0.2\pm 3.5  $ ($ -2.6$) \\ 
 \hline
 $b_1^{Sz1}$ &  $2.63^{+0.48}_{-0.32} $  ($2.28 $) & $1.39^{+0.15}_{-0.72}  $  ($1.05 $) &  $2.04\pm 0.14 $  ($ 2.09$) & $1.99^{+0.25}_{-0.20}  $ ($1.75 $) \\ 
 \hline
 $b_2^{Sz1}$ &  $-0.1^{+2.6}_{-3.0} $   ($-0.2 $) & $-1.3\pm 8.0$ ($ 2.91$) &  $0.2^{+2.3}_{-3.0}$  ($ -0.6$) & $1.1\pm 5.9 $ ($ -6.9$) \\ 
 \hline
 $b_{\mathcal{G}_2}^{Sz1}$ &  $0.4\pm 1.5 $  ($ 0.0$) & $-6.1\pm 5.0 $  ($-6.8 $) &  $0.3\pm 1.2$  ($ 0.7$) & $1.0^{+3.2}_{-2.8} $ ($ -2.3$) \\ 
 \hline
 $b_1^{Nz3}$ &  $2.71^{+0.69}_{-0.45} $  ($2.19 $) & $1.21^{+0.13}_{-0.73}  $  ($0.77 $) &  $1.88\pm 0.14$  ($1.90 $) & $1.85^{+0.22}_{-0.19}  $ ($1.70 $) \\ 
 \hline
 $b_2^{Nz3}$ &  $-4.7^{+2.6}_{-3.5}$  ($ -5.0$) & $-6.0\pm 7.8 $ ($-2.7 $) &  $-4.8^{+1.4}_{-2.0}$  ($ -4.1$) & $-4.9^{+1.7}_{-4.7} $ ($-7.4 $) \\ 
 \hline
 $b_{\mathcal{G}_2}^{Nz3}$ &  $-1.4\pm 1.8$  ($ -0.8$) & $-11.8\pm 5.5$  ($-12.9 $) &  $-0.9\pm 1.3  $  ($-1.3 $) & $-0.9^{+3.2}_{-2.8} $ ($ -2.5 $) \\ 
 \hline
 $b_1^{Sz3}$ &  $3.02^{+0.61}_{-0.42}   $ ($2.54 $) & $1.61^{+0.21}_{-0.85}$  ($1.29 $) &  $2.24\pm 0.13$  ($ 2.29$) & $2.23\pm 0.22  $ ($ 2.24 $) \\ 
 \hline
 $b_2^{Sz3}$ &  $-0.2\pm 2.9 $ ($ 0.0$) & $-2.1^{+8.8}_{-9.9}$ ($ -1.7$) &  $-0.4^{+2.3}_{-3.3} $  ($ -1.1$) & $-2.8^{+2.7}_{-7.7}  $ ($ -6.6$) \\ 
 \hline
 $b_{\mathcal{G}_2}^{Sz3}$ &  $-0.8\pm 1.7$  ($ 0.8$) & $-7.0\pm 5.8 $  ($ -10.5$) &  $0.0\pm 1.3 $  ($ -0.5$) & $1.6^{+3.4}_{-3.0} $ ($ 3.4$) \\ 
 \hline
\end{tabular}
\caption{Parameter constraints for all analyses with broader priors of nuisance parameters for $wA$CDM. We show results using full shape + BAO data both for CMB-free analyses and those with a Planck prior (labeled CMB). We show the 7 sampled cosmological parameters as well as 3 bias parameters per redshift bin and sky cut. The values shown are the means with the $1\sigma$ confidence intervals with the best-fit values in parenthesis.}
\label{tab_prior_wACDM_bias}
\end{table}

As seen previously, when the priors are broadened, large second-order biases are measured, particularly in the higher redshift bin, where $O(-5)$ or larger values are preferred for $b^{Nz3}_2$ in all cases. In the CMB-free case with a factor of 10 increase of the priors, even larger values are seen for various bias parameters, similar to those measured in the $\Lambda$CDM and $w$CDM versions of this exercise.

It would be interesting to evaluate the effect of priors in other beyond-$\Lambda$CDM full-shape analyses, such as the case of curvature~\cite{Glanville:2022xes} and modified gravity~\cite{Piga:2022mge}, particularly since those effects also modify the late-time amplitude.

\section{Full contours}
\label{sec:full_cont}

In this appendix, we show the contours for the full set of sampled parameters in each analysis, including all nuisance parameters that were not marginalised over analytically.

\begin{table}[h!]
\centering
 \begin{tabular}{||l l l ||} 
 \hline 
 Parameter & $\Lambda$CDM CMB prior & $\Lambda$CDM CMB-free \\
 \hline\hline
$\omega_c$ & $0.118\pm0.005$ ($0.120$) & $0.121^{+0.008}_{-0.009} $ ($0.120$) \\  
 \hline
 $100 \omega_b$ & $2.264\pm 0.037$ ($2.256$) & $2.266\pm 0.038 $ ($2.244$)\\ 
 \hline
 $h$ & $0.678\pm 0.008 $ ($0.681$) & $0.681\pm 0.010$ ($0.678$)\\ 
 \hline
$\log(10^{10} A_s)$ & $3.021\pm 0.040 $ ($3.023$)  & $2.821\pm 0.158 $ ($2.832$)\\  
 \hline
 $n_s$ & $0.969\pm 0.012 $ ($0.974$) & $0.975\pm 0.063$ ($0.972$) \\ 
 \hline\hline
 $b_1^{Nz1}$ &  $1.941\pm 0.068 $ ($1.908$) & $2.17\pm 0.15$ ($2.14$)\\ 
 \hline
 $b_2^{Nz1}$ &  $-0.71^{+0.77}_{-0.89} $ ($-1.10$) & $-0.35\pm 0.88$ ($-0.62$)\\ 
 \hline
 $b_{\mathcal{G}_2}^{Nz1}$ &  $-0.21\pm 0.39$ ($-0.2539$) & $-0.26\pm 0.43 $ ($-0.32$)\\ 
 \hline
 $b_1^{Sz1}$ &  $1.972\pm 0.076 $ ($1.96$) & $2.19\pm 0.15  $ ($2.16$)\\ 
 \hline
 $b_2^{Sz1}$ &  $-0.26^{+0.83}_{-0.94} $ ($0.04$) & $-0.03\pm 0.91$ ($-0.36$)\\ 
 \hline
 $b_{\mathcal{G}_2}^{Sz1}$ &  $0.16\pm 0.40$ ($0.27$) & $0.22\pm 0.44 $ ($0.00$)\\ 
 \hline
 $b_1^{Nz3}$ &  $1.988\pm 0.074 $ ($1.986$) & $2.25\pm 0.17  $ ($2.23$)\\ 
 \hline
 $b_2^{Nz3}$ &  $-1.90^{+0.88}_{-0.99}$ ($-1.73$) & $-1.3\pm 1.0$ ($-1.3$)\\ 
 \hline
 $b_{\mathcal{G}_2}^{Nz3}$ &  $-0.09\pm 0.44 $ ($-0.03$) & $-0.04\pm 0.50 $ ($-0.11$)\\ 
 \hline
 $b_1^{Sz3}$ &  $2.181\pm 0.079 $ ($2.186$) & $2.43\pm 0.17$ ($2.41$)\\ 
 \hline
 $b_2^{Sz3}$ &  $-0.17\pm 0.93$ ($-0.21$)& $0.01\pm 0.95 $ ($-0.01$)\\ 
 \hline
 $b_{\mathcal{G}_2}^{Sz3}$ &  $-0.11\pm 0.45$ ($-0.20$) & $-0.23\pm 0.52 $ ($-0.30$)\\ 
 \hline
\end{tabular}
\caption{Parameter constraints for the two $\Lambda$CDM analyses using full shape data as well as measurements of the BAO scale (FS+BAO). We show the 5 sampled cosmological parameters as well as 3 bias parameters per redshift bin and sky cut. The values shown are the means with the $1\sigma$ confidence intervals and the best-fit value in parenthesis.}
\label{tab_all_lcdm}
\end{table}

\begin{figure}[h!]
\centering 
\includegraphics[width=\textwidth]{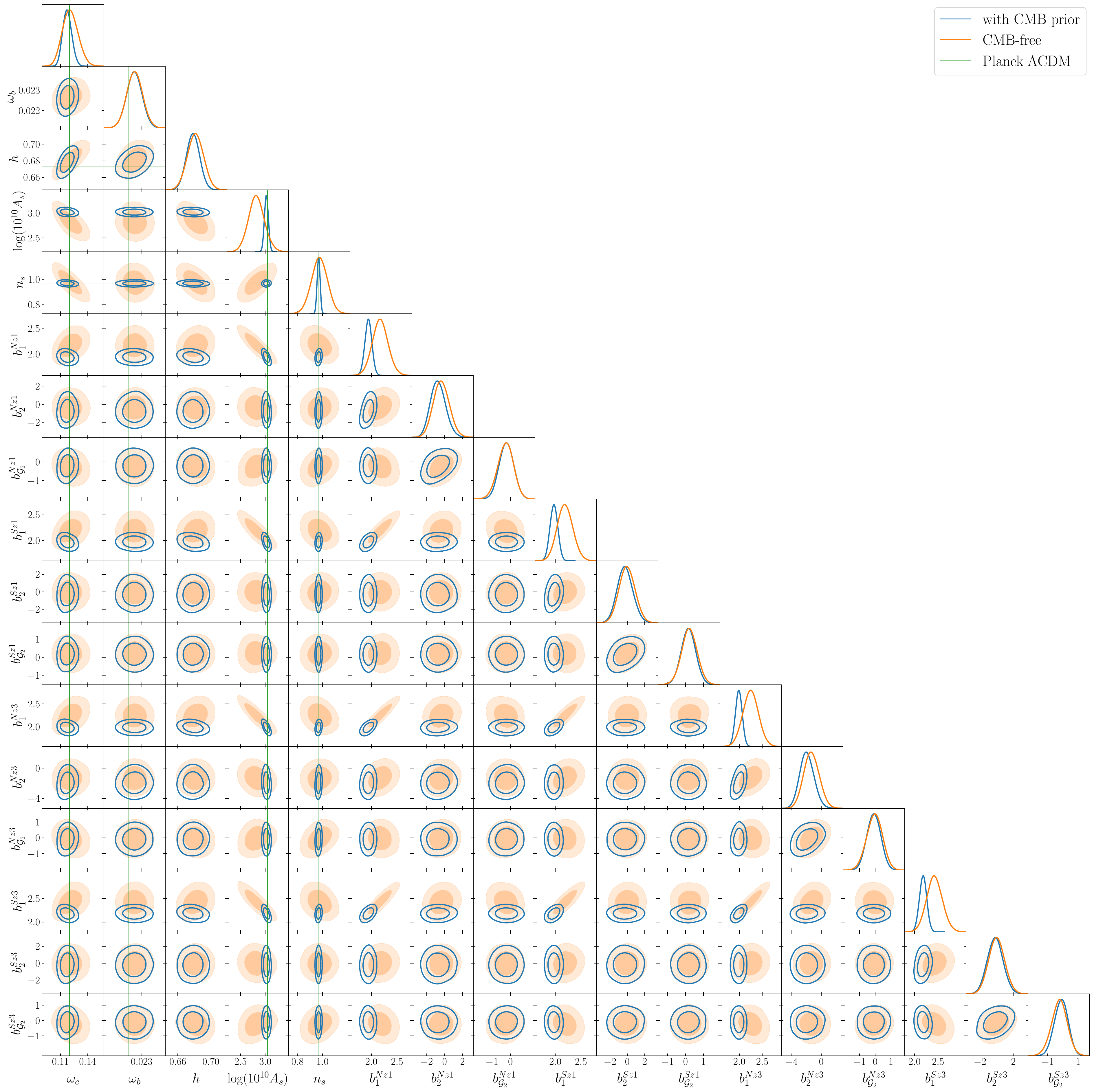}
\caption{\label{fig:LCDM_all_pars} Posteriors for all sampled parameters in the baseline analyses in $\Lambda$CDM, showing both the case with a CMB prior and without.}
\end{figure}

\begin{table}[h!]
\centering
 \begin{tabular}{||l l l ||} 
 \hline 
Parameter & $w$CDM CMB prior & $w$CDM CMB-free \\
 \hline\hline
 $w$ & $-1.002^{+0.081}_{-0.073}$ ($-1.023$) & $-1.17^{+0.12}_{-0.11} $ ($-1.093$) \\  
 \hline
 $\omega_c$ & $0.118\pm 0.005 $ ($0.118$) & $0.128^{+0.009}_{-0.011} $ ($0.122$) \\  
 \hline
 $100 \omega_b$ & $2.265\pm 0.038$ ($2.286$) & $2.269\pm 0.038  $ ($2.273$)\\ 
 \hline
 $h$ & $0.679^{+0.014}_{-0.016}$ ($0.689$) & $0.713\pm 0.024$ ($0.699$)\\ 
 \hline
$\log(10^{10} A_s)$ & $3.021\pm 0.041  $ ($3.027$)  & $2.62\pm 0.21 $ ($2.77$)\\  
 \hline
 $n_s$ & $0.968\pm 0.012 $ ($0.966$) & $0.947\pm 0.064$ ($0.9721$) \\ 
 \hline\hline
 $b_1^{Nz1}$ &  $1.942\pm 0.073 $ ($1.91$) & $2.31\pm 0.18$ ($2.19$)\\ 
 \hline
 $b_2^{Nz1}$ &  $-0.72^{+0.73}_{-0.89} $ ($-0.84$) & $-0.39\pm 0.90$ ($-0.40$)\\ 
 \hline
 $b_{\mathcal{G}_2}^{Nz1}$ &  $-0.19\pm 0.39 $ ($-0.2539$) & $-0.22\pm 0.44 $ ($-0.20$)\\ 
 \hline
 $b_1^{Sz1}$ &  $1.974\pm 0.080 $ ($1.96$) & $2.32\pm 0.18 $ ($2.23$)\\ 
 \hline
 $b_2^{Sz1}$ &  $-0.22\pm 0.89 $ ($-0.11$) & $-0.11\pm 0.91$ ($-0.34$)\\ 
 \hline
 $b_{\mathcal{G}_2}^{Sz1}$ &  $0.17\pm 0.40 $ ($0.00$) & $0.32\pm 0.46 $ ($0.00$)\\ 
 \hline
 $b_1^{Nz3}$ &  $1.987\pm 0.076$ ($1.960$) & $2.43\pm 0.22 $ ($2.297$)\\ 
 \hline
 $b_2^{Nz3}$ &  $-1.91^{+0.86}_{-1.0}$ ($-1.88$) & $-1.2\pm 1.0$ ($-1.5$)\\ 
 \hline
 $b_{\mathcal{G}_2}^{Nz3}$ &  $-0.10\pm 0.44 $ ($-0.07$) & $0.14\pm 0.53 $ ($-0.15$)\\ 
 \hline
 $b_1^{Sz3}$ &  $2.179\pm 0.082 $ ($2.168$) & $2.60\pm 0.22$ ($2.47$)\\ 
 \hline
 $b_2^{Sz3}$ &  $-0.17\pm 0.973$ ($-0.01$)& $0.00\pm 0.95 $ ($-0.01$)\\ 
 \hline
 $b_{\mathcal{G}_2}^{Sz3}$ &  $-0.12\pm 0.46$ ($-0.07$) & $-0.16\pm 0.55 $ ($-0.09$)\\ 
 \hline
\end{tabular}
\caption{Parameter constraints for the two $w$CDM analyses using full shape data as well as measurements of the BAO scale (FS+BAO). We show the 6 sampled cosmological parameters as well as 3 bias parameters per redshift bin and sky cut. The values shown are the means with the $1\sigma$ confidence intervals and the best-fit value in parenthesis.}
\label{tab_all_wcdm}
\end{table}

\begin{figure}[h!]
\centering 
\includegraphics[width=\textwidth]{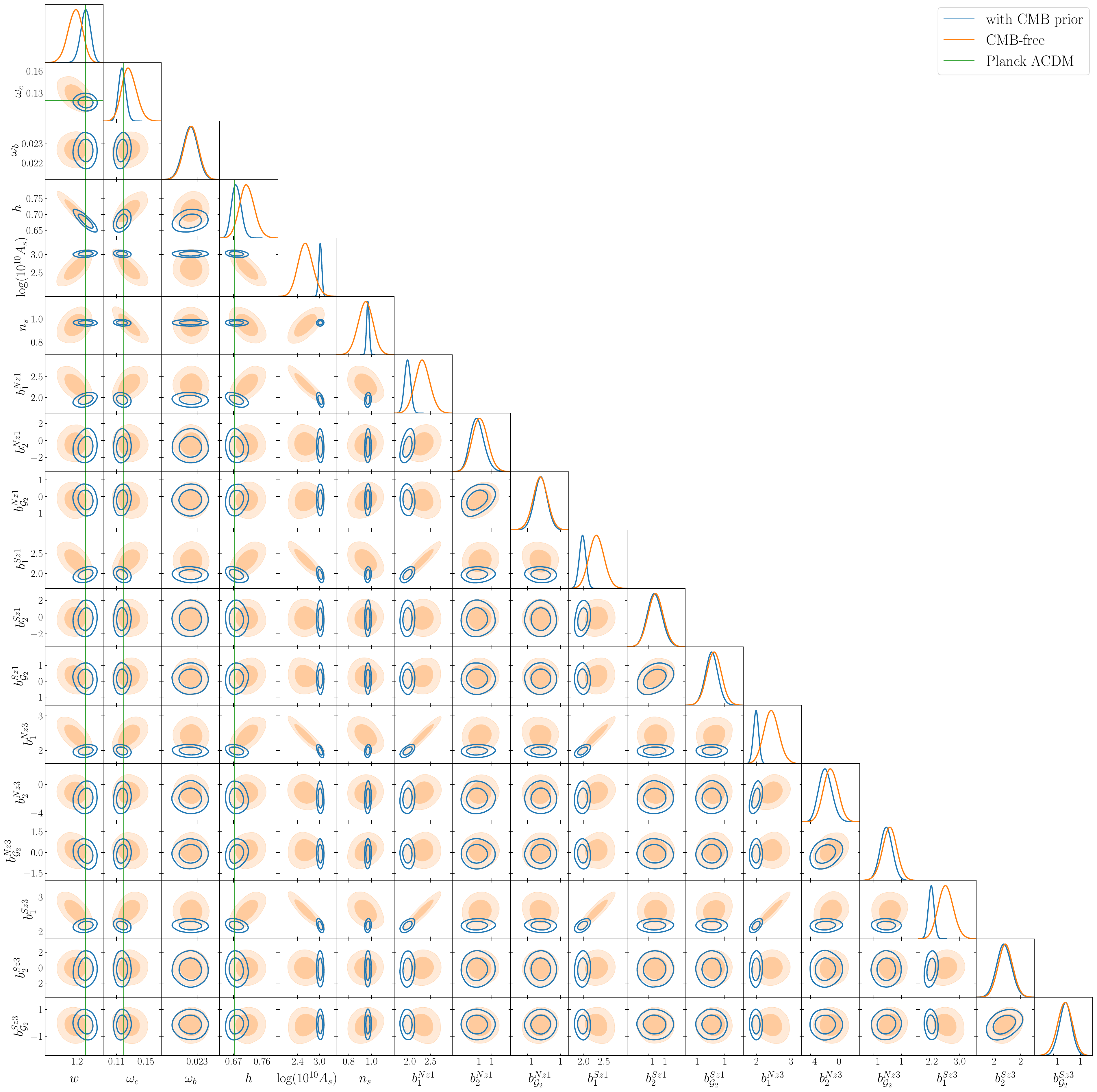}
\caption{\label{fig:wCDM_all_pars} Posteriors for all sampled parameters in the baseline analyses in $w$CDM, showing both the case with a CMB prior and without.}
\end{figure}

\begin{figure}[h]
\centering 
\includegraphics[width=\textwidth]{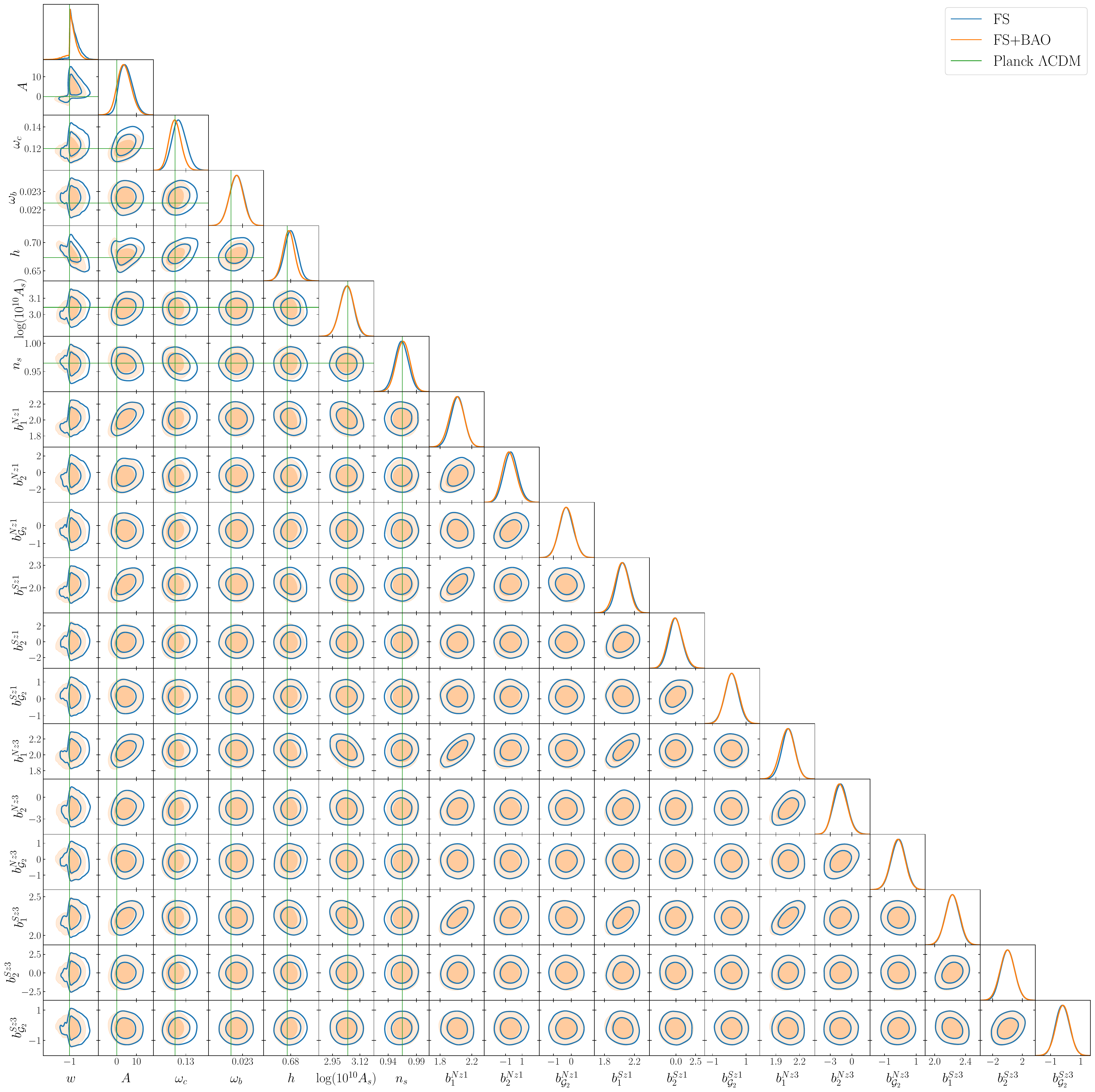}
\caption{\label{fig:baseline_all_pars} Posteriors for all sampled parameters in the CMB-dependent analyses in $wA$CDM, showing both results with FS data only and those with FS+BAO.}
\end{figure}

\begin{table}[h!]
\centering
 \begin{tabular}{||l l l ||} 
 \hline 
 Parameter & $wA$CDM FS & $wA$CDM FS+BAO \\
 \hline\hline
 $w$ &  $-0.954^{+0.024}_{-0.046}$ ($-0.987$) & $-0.972^{+0.036}_{-0.029} $ ($-0.994$) \\ 
 \hline
 $A$ &  $4.8^{+2.8}_{-3.8} $ ($3.7$) & $3.9^{+3.2}_{-3.7} $ ($5.4$)\\ \hline
$\omega_c$ & $0.1239^{+0.0062}_{-0.0069}$ ($0.1335$) & $0.1201^{+0.0052}_{-0.0058} $ ($0.1200$) \\  
 \hline
 $100 \omega_b$ & $2.267\pm 0.038$ ($2.296$) & $2.266\pm 0.038$ ($2.268$)\\ 
 \hline
 $h$ & $0.680\pm 0.012 $ ($0.6912$) & $0.677\pm 0.011$ ($0.677$)\\ 
 \hline
$\log(10^{10} A_s)$ & $3.038\pm 0.042$ ($3.060$)  & $3.039\pm 0.043$ ($3.033$)\\  
 \hline
 $n_s$ & $0.964\pm 0.012 $ ($0.958$) & $0.966\pm 0.012$ ($0.962$) \\ 
 \hline\hline
 $b_1^{Nz1}$ &  $2.020\pm 0.085$ ($1.953$) & $2.011\pm 0.090$ ($2.030$) \\ 
 \hline
 $b_2^{Nz1}$ &  $-0.36\pm 0.84$ ($-0.55$) & $-0.53^{+0.79}_{-0.88}$ ($-0.48$)\\ 
 \hline
 $b_{\mathcal{G}_2}^{Nz1}$ &  $-0.28\pm 0.40$ ($-0.486$) & $-0.28\pm 0.39 $ ($-0.29$)\\ 
 \hline
 $b_1^{Sz1}$ &  $2.047\pm 0.090 $ ($ 1.974$) & $2.033\pm 0.092 $ ($2.064$) \\ 
 \hline
 $b_2^{Sz1}$ &  $-0.04\pm 0.85 $ ($-0.52$) & $-0.07^{+0.86}_{-0.95}$ ($-0.38$)\\ 
 \hline
 $b_{\mathcal{G}_2}^{Sz1}$ &  $0.14\pm 0.41 $ ($-0.32$) & $0.16\pm 0.41  $ ($0.00$)\\ 
 \hline
 $b_1^{Nz3}$ &  $2.059\pm 0.085$ ($1.963$) & $2.045\pm 0.090$ ($2.065$) \\ 
 \hline
 $b_2^{Nz3}$ &  $-1.55\pm 0.97$ ($-2.10$) & $-1.62^{+0.90}_{-1.0}$ ($-1.81$)\\ 
 \hline
 $b_{\mathcal{G}_2}^{Nz3}$ &  $-0.12\pm 0.45$ ($-0.08$) & $-0.16\pm 0.45 $ ($-0.29$)\\ 
 \hline
 $b_1^{Sz3}$ &  $2.229\pm 0.090$ ($2.104$) & $2.232\pm 0.093$ ($2.232$) \\ 
 \hline
 $b_2^{Sz3}$ &  $0.04\pm 0.93$ ($0.83$) & $-0.01\pm 0.97$ ($0.00$)\\ 
 \hline
 $b_{\mathcal{G}_2}^{Sz3}$ &  $-0.20\pm 0.45$ ($0.20$) & $-0.19\pm 0.46$ ($-0.25$)\\ 
 \hline
\end{tabular}
\caption{Parameter constraints for the two CMB-dependent analyses for $wA$CDM. We show results using full shape only (FS) as well those including measurements of the BAO scale (FS+BAO). We show the 7 sampled cosmological parameters as well as 3 bias parameters per redshift bin and sky cut. The values shown are the means with the $1\sigma$ confidence intervals and the best-fit value in parenthesis.}
\label{tab_baseline_BAO_bias}
\end{table}

\begin{figure}[h!]
\centering 
\includegraphics[width=\textwidth]{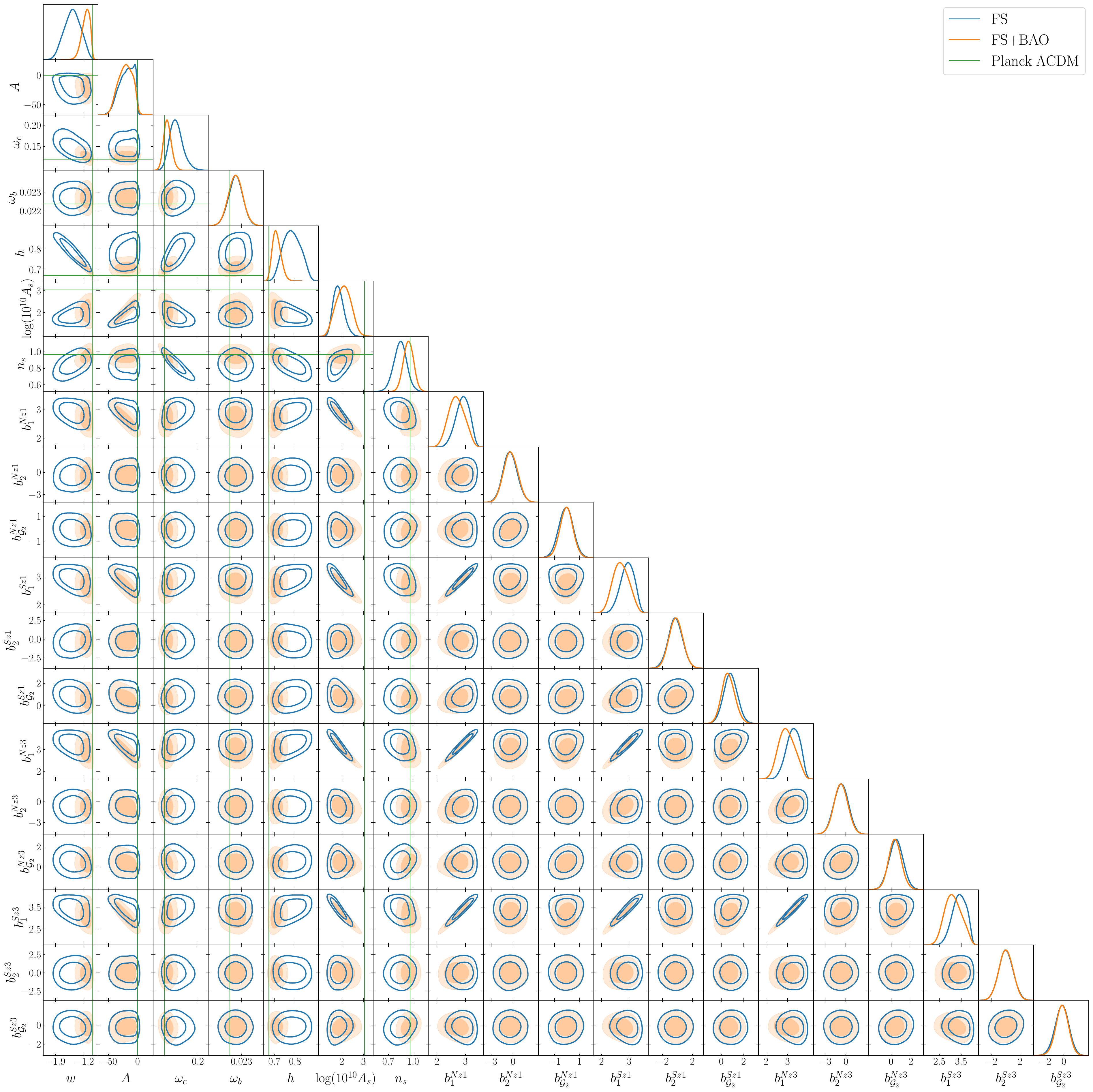}
\caption{\label{fig:cmb_free_all_pars} Posteriors for all sampled parameters in the CMB-free analyses in $wA$CDM, showing both results with FS data only and those with FS+BAO.}
\end{figure}

\begin{table}[h!]
\centering
 \begin{tabular}{||l l l ||} 
 \hline 
 Parameter & $wA$CDM FS & $wA$CDM FS+BAO \\
 \hline\hline
$w$ &  $-1.48\pm 0.20$ ($-1.28$) & $-1.16^{+0.12}_{-0.069} $ ($-1.097$) \\ 
 \hline
 $A$ &  $-18.9^{+17}_{-7.7}  $ ($-0.7631$) & $-21^{+14}_{-11}  $ ($-0.1077$)\\ 
 \hline
$\omega_c$ & $0.148^{+0.013}_{-0.018}$ ($0.138$) & $0.1275^{+0.0082}_{-0.010} $ ($0.1249$) \\  
 \hline
 $100 \omega_b$ & $2.269\pm 0.038$ ($2.266$) & $2.268\pm 0.038        $ ($2.253$)\\ 
 \hline
 $h$ & $0.785^{+0.041}_{-0.046}$ ($0.746$) & $0.711^{+0.017}_{-0.024}$ ($ 0.6992$)\\ 
 \hline
$\log(10^{10} A_s)$ & $1.85^{+0.20}_{-0.28} $ ($2.37$)  & $2.07\pm 0.35$ ($2.725$)\\  
 \hline
 $n_s$ & $0.844\pm 0.080$ ($0.885$) & $0.945\pm 0.062$ ($0.9553$) \\ 
 \hline\hline
 $b_1^{Nz1}$ &  $2.90^{+0.27}_{-0.21}$ ($2.45$) & $2.69\pm 0.28$ ($2.215$) \\ 
 \hline
 $b_2^{Nz1}$ &  $-0.40\pm 0.97$ ($-0.63$) & $-0.34\pm 0.96$ ($-0.66$)\\ 
 \hline
 $b_{\mathcal{G}_2}^{Nz1}$ &  $-0.09\pm 0.57$ ($-0.33$) & $-0.05\pm 0.52$ ($-0.25$)\\ 
 \hline
 $b_1^{Sz1}$ &  $2.94^{+0.28}_{-0.22}$ ($2.47$) & $2.71\pm 0.29$ ($2.245$) \\ 
 \hline
 $b_2^{Sz1}$ &  $-0.27\pm 0.95$ ($-0.39$) & $-0.21\pm 0.92$ ($-0.28$)\\ 
 \hline
 $b_{\mathcal{G}_2}^{Sz1}$ &  $0.83\pm 0.62 $ ($0.00$) & $0.64^{+0.55}_{-0.62}$ ($0.00$)\\ 
 \hline
 $b_1^{Nz3}$ &  $3.26^{+0.36}_{-0.28}$ ($2.65$) & $2.97^{+0.36}_{-0.42}$ ($2.32$) \\ 
 \hline
 $b_2^{Nz3}$ &  $-0.7\pm 1.0$ ($-1.2$) & $-0.7\pm 1.0$ ($-1.6$)\\ 
 \hline
 $b_{\mathcal{G}_2}^{Nz3}$ &  $0.51\pm 0.71 $ ($-0.23$) & $0.40\pm 0.65$ ($-0.15$)\\ 
 \hline
 $b_1^{Sz3}$ &  $3.42^{+0.37}_{-0.28}$ ($2.77$) & $3.14^{+0.36}_{-0.43}$ ($2.51$) \\ 
 \hline
 $b_2^{Sz3}$ &  $0.03\pm 0.99$ ($0.00$) & $0.03\pm 0.97$ ($0.00$)\\ 
 \hline
 $b_{\mathcal{G}_2}^{Sz3}$ &  $-0.20\pm 0.73$ ($-0.25$) & $-0.19\pm 0.68$ ($-0.28$)\\ 
 \hline
\end{tabular}
\caption{Parameter constraints for the two CMB-free analyses for $wA$CDM. We show results using full shape only (FS) as well those including measurements of the BAO scale (FS+BAO). We show the 7 sampled cosmological parameters as well as 3 bias parameters per redshift bin and sky cut. The values shown are the means with the $1\sigma$ confidence intervals and the best-fit value in parenthesis.}
\label{tab_cmb_free_BAO_bias}
\end{table}

\end{document}